\documentclass[fleqn,usenatbib]{mnras}

\usepackage{newtxtext,newtxmath}
 
\usepackage[T1]{fontenc}
\usepackage{ae,aecompl}

\usepackage{graphicx}	
\usepackage{amsmath}	
\usepackage{amssymb}	
\usepackage{makecell}

\title[AGN properties by CIGALE SED fitting]{Extinction-free Census of AGNs in the $AKARI$/IRC North Ecliptic Pole Field from 23-band Infrared Photometry from Space Telescopes}

\author[Ting-Wen Wang et al.]{
Ting-Wen Wang$^{1}$\thanks{E-mail: tinattw0127@gapp.nthu.edu.tw},
Tomotsugu Goto$^{1}$,
Seong Jin Kim$^{1}$,
Tetsuya Hashimoto$^{1,2}$,
\newauthor
Denis Burgarella$^{3}$,
Yoshiki Toba$^{4,5,6}$,
Hyunjin Shim$^{7}$,
Takamitsu Miyaji$^{8,9}$\footnote{On sabbatical leave from IA-UNAM-E at AIP.},
\newauthor
Ho Seong Hwang$^{10}$,
Woong-Seob Jeong$^{10,11}$,
Eunbin Kim$^{10}$,
Hiroyuki Ikeda$^{12}$,
\newauthor
Chris Pearson$^{13,14,15}$,
Matthew Malkan$^{16}$,
Nagisa Oi$^{17}$,
Daryl Joe D. Santos$^{1}$,
\newauthor
Katarzyna Ma\l{}ek$^{3,18}$,
Agnieszka Pollo$^{18,19}$
Simon C.-C. Ho$^{1}$,
Hideo Matsuhara$^{20,21}$,
\newauthor
Alvina Y. L. On$^{1,2,22}$,
Helen K. Kim$^{16}$,
Tiger Yu-Yang Hsiao$^{5,23}$,
and Ting-Chi Huang$^{20,21}$
\\
$^{1}$Institute of Astronomy, National Tsing Hua University, No. 101, Section 2, Kuang-Fu Road, Hsinchu City 30013, Taiwan\\
$^{2}$Centre for Informatics and Computation in Astronomy (CICA), National Tsing Hua University, 101, Section 2. Kuang-Fu Road,\\ Hsinchu, 30013, Taiwan (R.O.C.)\\
$^{3}$Aix Marseille Univ. CNRS, CNES, LAM Marseille, France\\
$^{4}$Department of Astronomy, Kyoto University, Kitashirakawa-Oiwake-cho, Sakyo-ku, Kyoto 606-8502, Japan\\
$^{5}$Academia Sinica Institute of Astronomy and Astrophysics, 11F of Astronomy-Mathematics Building, AS/NTU \\
No.1, Section 4, Roosevelt Road, Taipei 10617, Taiwan\\
$^{6}$Research Center for Space and Cosmic Evolution, Ehime University, 2-5 Bunkyo-cho, Matsuyama, Ehime 790-8577, Japan\\
$^{7}$Department of Earth Science Education, Kyungpook National University, 80 Daehak-ro, Buk-gu, Daegu 41566, Republic of Korea\\
$^{8}$Instituto de Astrnom\'ia sede Ensenada, Universidad Nacinal Aut\'onoma de M\'exico (IA-UNAM-E) Km 107, Carret. Tij.-Ens., 22860,\\
Ensenada,BC, Mexico\\
$^{9}$Leibnitz Instituto f\"ur Astrophysik (AIP), An der Sternwarte 16, 14482, Potsdam, Germany \\
$^{10}$Korea Astronomy and Space Science Institute, 776 Daedeokdae-ro, Yuseong-gu, Daejeon 34055, Republic of Korea\\
$^{11}$Korea University of Science and Technology, 217 Gajeong-ro, Yuseong-gu, Daejeon 34113, Republic of Korea\\
$^{12}$National Institute of Technology, Wakayama College, Gobo, Wakayama 644-0023, Japan\\
$^{13}$RAL Space, STFC Rutherford Appleton Laboratory, Didcot, Oxon, OX11 0QX, UK\\
$^{14}$The Open University, Milton Keynes, MK7 6AA, UK\\
$^{15}$University of Oxford, Keble Rd, Oxford, OX1 3RH, UK\\
$^{16}$Department of Physics and Astronomy, UCLA, 475 Portola Plaza, Los Angeles, CA 90095-1547, USA\\
$^{17}$Tokyo University of Science, 1-3, Kagurazaka Shinjuku-ku Tokyo 162-8601 Japan\\
$^{18}$National Centre for Nuclear Research, ul. Pasteura 7, 02-931 Warszawa\\
$^{19}$Astronomical Observatory of the Jagiellonian University, ul.Orla 171, 30-244 Krakow, Poland\\
$^{20}$Department of Space and Astronautical Science, Graduate University for Advanced Studies, SOKENDAI, Shonankokusaimura, \\Hayama, Miura District, Kanagawa 240-0193, Japan\\
$^{21}$Institute of Space and Astronautical Science, Japan Aerospace Exploration Agency, 3-1-1 Yoshinodai, Chuo-ku, Sagamihara,\\Kanagawa 252-5210, Japan\\
$^{22}$Mullard Space Science Laboratory, University College London, Holmbury St. Mary, Dorking, Surrey, RH5 6NT, United Kingdom \\ 
$^{23}$Department of Atmospheric Science, National Central University, No.300, Zhongda Rd., Zhongli Dist., Taoyuan City 32001, Taiwan (R.O.C.)\\
}

\date{Accepted XXX. Received YYY; in original form ZZZ}

\pubyear{2020}

\begin{document}
\label{firstpage}
\pagerange{\pageref{firstpage}--\pageref{lastpage}}
\maketitle
\clearpage
\begin{abstract}

In order to understand the interaction between the central black hole and the whole galaxy or their co-evolution history along with cosmic time, a complete census of active galactic nuclei (AGN) is crucial. However, AGNs are often missed in optical, UV and soft X-ray observations since they could be obscured by gas and dust. A mid-infrared (mid-IR) survey supported by multiwavelength data is one of the best ways to find obscured AGN activities because it suffers less from extinction. Previous large IR photometric surveys, e.g., $WISE$ and $Spitzer$, have gaps between the mid-IR filters. Therefore, star forming galaxy (SFG)-AGN diagnostics in the mid-IR were limited. The $AKARI$ satellite has a unique continuous 9-band filter coverage in the near to mid-IR wavelengths. In this work, we take advantage of the state-of-the-art spectral energy distribution (SED) modelling software, \begin{footnotesize}CIGALE\end{footnotesize}, to find AGNs in mid-IR. We found 126 AGNs in the NEP-Wide field with this method. We also investigate the energy released from the AGN as a fraction of the total IR luminosity of a galaxy. We found that the AGN contribution is larger at higher redshifts for a given IR luminosity. With the upcoming deep IR surveys, e.g., $JWST$, we expect to find more AGNs with our method.

\end{abstract}

\begin{keywords}
galaxies: active -- infrared: galaxies
\end{keywords}



\section{Introduction}

An active galactic nucleus (AGN) is a compact ultra-luminous region at the centre of a galaxy. It is widely believed that a supermassive black hole (SMBH) resides in every AGN, and the mass of a black hole is related to the bulge mass of a galaxy \citep[e.g., ][]{Kormendy2013}. In order to study galaxy evolution, it is important to find AGNs in the Universe, as the prescence of an AGN has a nonnegligible impact on the main physical parameters of galaxies, such as star formation rate (SFR), etc. 
There are several ways to search for AGNs, e.g., based on optical, UV and soft X-ray observations \citep[e.g., ][]{Hickox2018}. 
However, AGNs may be obscured by gas and dust. The obscured AGNs could be missed by these observations. 
Also, Compton-thick AGNs (CTAGNs) are highly obscured, even at hard X-rays with $E\ga 10$ keV.
To avoid missing dust-obscured AGNs in the Universe, mid-infrared (mid-IR) survey of AGNs is crucial.

Mid-IR surveys are sensitive to AGNs due to the warm dust emission of AGNs. In general, AGNs show thermal emission from warm dust at mid-IR wavelengths, which is heated by strong radiation from the central accretion disks. However, in the mid-IR wavelengths such as 3.3, 6.2, 7.7, 8.6, and 11.3 $\mu m$, star-forming galaxies (SFGs) also produce strong polycyclic aromatic hydrocarbon (PAH) emission \citep[e.g., ][]{Feltre2013, Ohyama2018, Kim2019}.
Therefore, in order to find dust-obscured AGNs using mid-IR surveys, we need to carefully examine the spectral features in mid-IR wavelengths. 

Many works have been conducted to search, identify and characterise AGNs by using the data from $Spitzer$ infrared telescope and Wide field Infrared Survey Explorer ($WISE$) \citep[e.g., ][]{Hwang2012, Toba2015, Toba2016}. However, both $WISE$ and $Spitzer$ have limited available filters (3.4, 4.6, 12 and 22 $\mu m$ in $WISE$ and 3.6, 4.5, 5.8, 8.0 and 24 $\mu m$ in $Spitzer$) and there are wavelength gaps within the mid-IR wavelengths such as 10-20 $\mu m$, making SFG-AGN diagnostics difficult.
If the feature of a source falls in the gap of the filters in $WISE$ or $Spitzer$, we may fail to classify SFG and AGN correctly.

To avoid this type of failure, the best way is to use an instrument with continuous filter system covering these mid-IR gaps, such as the Infrared Camera (IRC, \citet{Onaka2007}) installed in $AKARI$ space telescope. $AKARI$ \citep{Murakami2007}, an IR space telescope launched by ISAS/JAXA in 2006, successfully carried out an all-sky survey at IR wavelengths. The North Ecliptic Pole (NEP) survey \citep{Matsuhara2006} was one of the dedicated photometry surveys with the $AKARI$'s pointing observations. $AKARI$ observed the sky region with its 9 continuous passbands, from near- to mid-IR wavelengths. Detailed information on $AKARI$ is in Section~\ref{sec2}.

The combination of the $AKARI$ IR photometry and X-ray data gives a strong tool to find Compton-thick AGNs \citep{Krumpe2015,Miyaji2018}. \citet{Huang2017} performed spectral energy distribution (SED) fitting to select AGNs from data of the $AKARI$ NEP deep survey \citep{Matsuhara2006} covering a field of 0.57 deg$^2$, containing $\sim$ 5800 sources. By combining $AKARI$'s 9 passbands, $WISE$ ALLWISE 1, 2, 3, 4 and $Spitzer$ Infrared Array Camera (IRAC) 1, 2, 3, 4 and Multiband Imaging Photometer for $Spitzer$ (MIPS) 1 passbands, they fit 25 empirical models by using an SED template fitting code, \begin{footnotesize}Le Phare\end{footnotesize} \citep[PHotometric Analysis for Redshift Estimate;][]{Arnouts1999, Ilbert2006}. Because of the continuous mid-IR coverage of $AKARI$, they recovered more X-ray selected AGNs \citep{Krumpe2015} than previous works \citep{Lacy2007} based on IR colour-colour diagrams by $\sim$20 percent.

\citet{Chiang2019} also used \begin{footnotesize}Le Phare\end{footnotesize} to select AGNs by taking advantage of the unique $AKARI$ NEP wide field survey \citep[{NEP-Wide};][]{Matsuhara2006, Lee2009, Kim2012} sample with 18 IR bands of data, including $AKARI$'s 9 passbands, $WISE$ 1-4, $Spitzer$ IRAC 1-4 and MIPS 1 photometry. Their results indicate that AGN number fraction seems to show stronger IR luminosity dependence than redshift dependence. They also examined the fractions of SFGs and found mild decreasing trends at high IR luminosities.

Recently, Subaru Hyper Suprime-Cam \citep[{HSC};][]{Miyazaki2018} observations with $g$, $r$, $i$, $z$, and $Y$ bands were carried out to cover the entire field of the $AKARI$ NEP Wide field (covering 5.4 deg$^2$) \citep{Goto2017}. 
The new HSC observation allowed us to extend $AKARI$ sources with photometric redshift to the full coverage of the NEP Wide field \citep{Ho2020}. Furthermore, the advanced SED fitting code \begin{footnotesize}CIGALE\end{footnotesize}\footnote{\url{https://cigale.lam.fr}} \citep[Code Investigating GALaxy Emission;][]{Burgarella2005, Noll2009,Boquien2019} can not only classify AGNs but also obtain physical properties of our targets.

With the new HSC photometry and the state-of-the-art SED-fitting software, \begin{footnotesize}CIGALE\end{footnotesize}, our goal is to recover more dust-obscured AGNs in the $AKARI$ NEP-Wide survey and investigate the physical properties of the sources. 

This work is organised as follows. We describe our sample selection and method in Section~\ref{sec2}.
The result of our AGN selection and the physical properties of AGNs are described in Section~\ref{sec3}.
Our discussion is shown in Section~\ref{sec4}. Our conclusions are in Section~\ref{sec5}. Throughout this paper, we use AB magnitude system unless otherwise mentioned. We assume a cosmology with H$_0$ = 70.4 km s$^{-1}$Mpc$^{-1}$, {$\Omega_\Lambda$} = 0.728, and\ {$\Omega_\text{M}$} = 0.272 \citep{Komatsu2011}.

\section{Data and Analysis} \label{sec2}

\subsection{Data}

We selected samples from the $AKARI$ NEP Wide survey. The NEP Wide survey was centered at (RA, Dec)=(18${h}$00${m}$00${s}$, +66$^{\circ}$33${'}$38${''}$), covering 5.4 deg$^{2}$ with the $AKARI$ Infrared Camera (IRC). The $AKARI$ IRC has 3 channels: NIR, MIR-S and MIR-L. There are 3 filters in each channel, so the IRC has 9 filters in total: $N2$, $N3$, $N4$, $S7$, $S9W$, $S11$, $L15$, $L18W$, and $L24$, corresponding to 2, 3, 4, 7, 9, 11, 15, 18, and 24 $\mu$m of the reference wavelengths, respectively. \citet{Kim2012} presented a photometric catalogue of IR sources from the NEP-Wide survey using the nine photometric filters of the IRC. In the near-IR bands, the $N2$ filter reaches a depth of $\sim$20.9 mag, and the $N3$ and $N4$ bands reach $\sim$21.1 mag. The mid-IR detection limits are much shallower than those of near-IR bands: $\sim$19.5 ($S7$), 19.3 ($S9W$), and 19.0 ($S11$) for the MIR-S bands, and $\sim$18.6, 18.7, and 17.8, corresponding to $L15$, $L18W$ and $L24$ for the MIR-L bands, respectively.
Recently, Oi et al. (2020, submitted) cross-matched the Subaru Hyper Suprime-Cam (HSC) sources with the $AKARI$ sources in the NEP Wide field and constructed an HSC-$AKARI$ catalogue (see also \citet{Toba2020} for $AKARI$ sources without HSC counterparts). The HSC optical imaging observations were carried out in five broad bands ($g$, $r$, $i$, $z$, and $Y$) covering an almost entire field of the NEP-Wide survey ($\sim$5.4 deg$^{2}$). This catalogue enables us to calculate the photometric redshifts of the $AKARI$ sources.

In order to optimise the advantage of $AKARI$ photometry, \citet{Kim2020} constructed a multiwavelength catalogue based on $AKARI$ data. They used the ``matching radii'' determined by astrometry offset tests for each matching procedure. The multiwavelength catalogue contains UV to submillimetre counterparts of the $AKARI$ sources in the NEP-Wide field. In this catalogue, there are 91,861 sources in total. Among them, there are 2,026 sources with spectroscopic redshifts. Spectroscopic redshifts are collected from various published and unpublished sources as detailed below. The spectroscopic data are provided by several spectroscopic observations with various telescopes/instruments in optical,
Keck/DEIMOS \citep{Shogaki2018, Kim2018}, MMT/Hectospec \citep{Shim2013}, WIYN/Hydra \citep{Shim2013}, GTC \citep[Miyaji et al. in prep.;][]{Diaz-Tello2017, Krumpe2015}, and in near-IR, Subaru/FMOS \citep{Oi2017}. For the sources without spectroscopic redshifts, photometric reshifts of our sample are calculated by using \begin{footnotesize}Le Phare\end{footnotesize} \citep{Ho2020}. 
Also, \citet{Kim2020} obtained the X-ray catalogue from Chandra NEP deep survey and the spectra of 1,796 sources were obtained from \citep{Shim2013}. \citet{Shim2013} identified 1,128 star-forming or absorption-line-dominated galaxies, 198 Type-1 AGNs, 8 Type-2 AGNs, 121 Galactic stars, and 190 spectra of unknown sources due to low signal-to-noise ratio. Most of the sources with spectroscopic redshifts are already classified by \citet{Shim2013} and \citet{Krumpe2015}.

\begin{figure}
    \centering
	\includegraphics[width=\columnwidth]{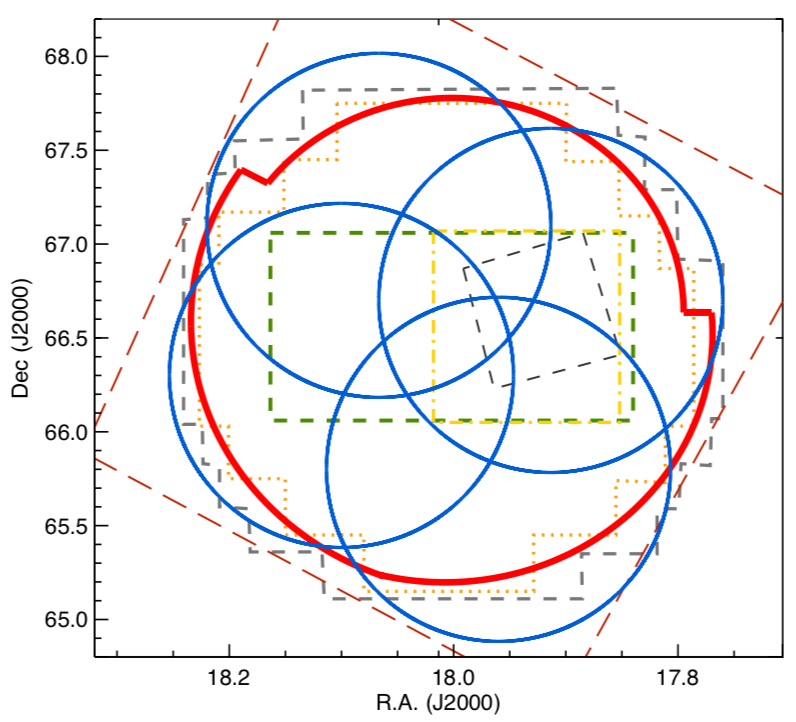}
    \caption{An overall map showing a variety of surveys around the NEP. The red circular area shows the $AKARI$'s NEP-Wide field \citep{Kim2012}. The four blue circles represent  the $r$-band coverage of the Subaru/HSC survey (Oi et al. 2020, submitted). The gray dashed line represents the optical surveys done with the the Maidanak SNUCAM ($B$, $R$, $I$) \citep{Jeon2010}. The green dashed rectangle represents the CFHT/MegaCam ($u^{\ast}$, $g$, $r$, $i$, $z$) survey \citep{Hwang2007}. The yellow square shows an observation with MegaCam and WIRCam ($Y$, $J$, $K_{s}$) on the NEP-Deep field \citep{Oi2014}. The orange dotted shape indicates the $H$ band survey with KPNO/FLAMINGOS \citep{Jeon2014}. The black dashed square inside the yellow box shows the area observed by the $Herschel$/PACS \citep{Pearson2019}. The largest dark-red rhombus shows the $Herschel$/SPIRE coverage (Pearson et al. in prep). The $g$, $i$, $z$ and $y$ band observations by the $Subaru$ HSC (Oi et al. 2020, submitted) is not shown in this figure. The $u$-band observation by the MegaPrime/MegaCam \citep{Huang2020} is not shown in this figure.}
    \label{fig:fig1}
\end{figure}

\begin{figure}
    \centering
	\includegraphics[width=\columnwidth]{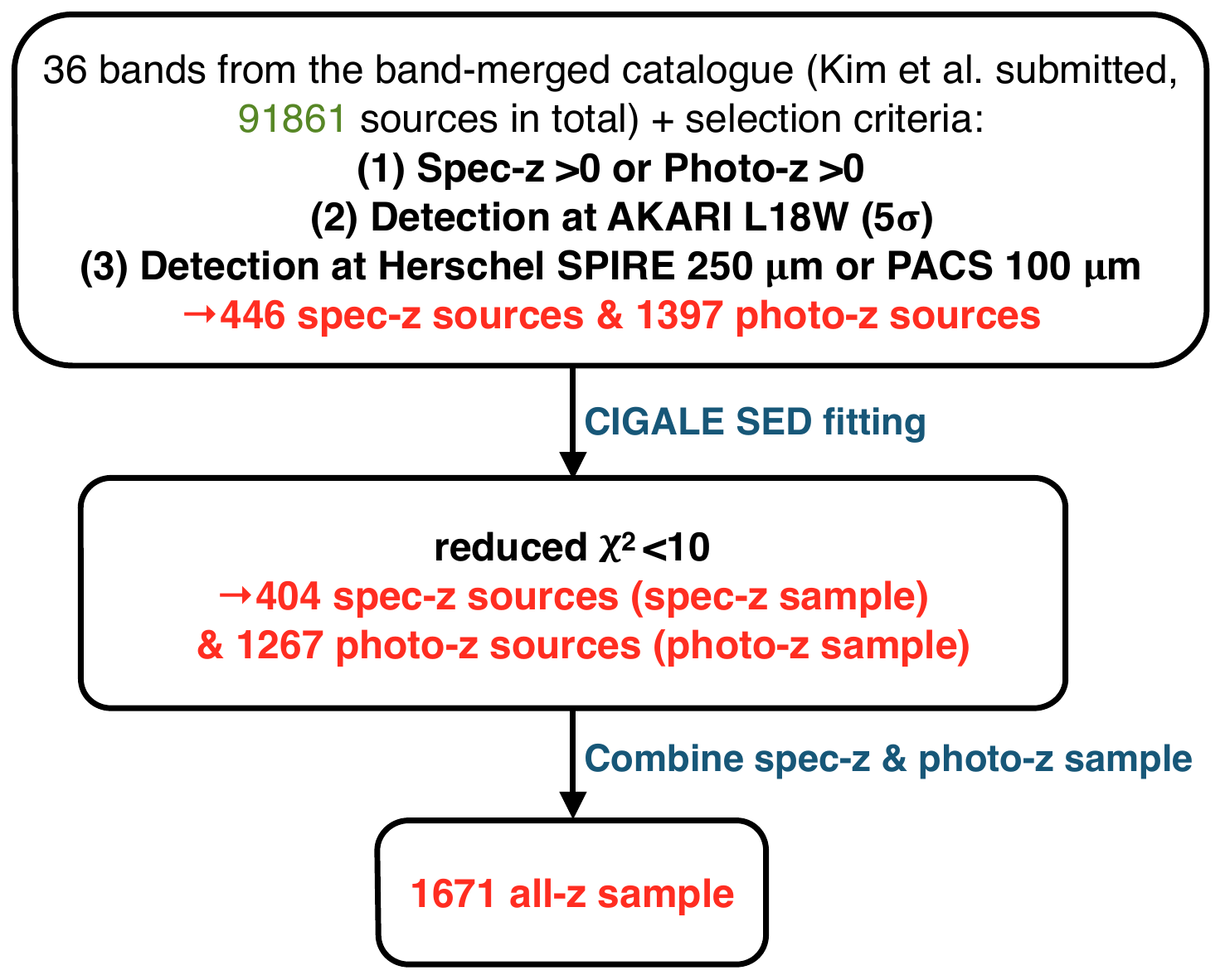}
    \caption{Flowchart of the sample selection process.}
    \label{fig:fig2}
\end{figure}

There are 42 bands in total in the multiwavelength catalogue \citep{Kim2020}. We used 36 bands from the catalogue in this work, including data in the UV band from CFHT MegaCam ($u^{*}$) and CFHT MegaPrime (u), 5 optical bands from Subaru HSC ($g$, $r$, $i$, $z$, $Y$), 3 bands from Maidanak SNUCAM ($B$, $R$, $I$), 1 band from FLAMINGOS ($H$), 9 bands from $AKARI$ IRC (2.4, 3.2, 4.1, 7, 9, 11, 15, 18, and 24 $\mu$m), 3 bands from CFHT WIRCam ($Y$, $J$, $K_s$), 4 mid-IR bands from $WISE$ ALLWISE 1-4 (3.4, 4.6, 12 and 22$\mu$m), 4 mid-IR bands from $Spitzer$ IRAC 1-4 (3.6, 4.5, 5.8 and 8.0$\mu$m), 2 bands from $Herschel$ PACS (100 and 160 $\mu$m), and 3 bands from $Herschel$ SPIRE (250, 350 and 500 $\mu$m).
An overall map showing a variety of surveys around the NEP is shown in Fig.~\ref{fig:fig1}. Information on the filters is given below.
Aside from the $AKARI$ IRC observational data, we also use other near to mid-IR data in this work.

The $H$-band catalogue \citep{Jeon2014} of the NEP region is obtained from the Florida Multi-object Imaging Near-IR Grism Observational Spectrometer \citep[FLAMINGOS; ][]{Elston2006} on the Kitt Peak National Observatory (KPNO).
KPNO has a 2.1 m telescope covering a 5.1 deg$^2$ area down to a 5$\sigma$ depth of $~$21.3 mag for the $H$-bands with an astrometric accuracy of 0.14 and 0.17 for 1$\sigma$ in the R.A. and Dec. directions, respectively.

We obtain the $WISE$ 4-band data from \citet{Jarrett2011}. $WISE$ is a NASA-funded Medium-Class Explorer mission, consisting of a 40-cm space-based infrared telescope whose science payload consists of mega-pixel cameras, cooled with a two-stage solid hydrogen cryostat. $WISE$ mapped the entire sky at 3.4$\mu$m (W1), 4.6$\mu$m (W2), 12$\mu$m (W3), and 22 $\mu$m (W4) with 5$\sigma$ depth of 18.1, 17.2, 18.4, and 16.1 mag, respectively.

The CFHT WIRCam data are obtained from \citet{Oi2014}. WIRCam uses four 2048 $\times$ 2048 HAWAII2RG CCD arrays. The 5$\sigma$ limiting magnitudes are 23.4, 23.0, and 22.7 for WIRCam $Y$, $J$, and $K_s$-bands, respectively.

The $Spitzer$ two-band catalogue of the NEP field IRAC1 (3.6 $\mu$m) and IRAC2 (4.5 $\mu$m) is presented by \citet{Nayyeri2018}. The observations covered 7 deg$^2$, with the 5$\sigma$ depths of 21.8 and 22.4 mag in the IRAC1 and IRAC2 bands, respectively.
The IRAC3(5.8 $\mu$m) and IRAC4(8 $\mu$m) data are obtained from \citet{Jarrett2011}. The observations covered 0.40 deg$^2$, with the 5$\sigma$ depths of 16.6 and 15.4 mag in the IRAC3 and IRAC4 bands, respectively.

We used the data from Subaru HSC \citep[][ Oi et al. 2020, submitted]{Goto2017}. The deep HSC $g$, $r$, $i$, $z$ and $Y$ imaging of the $AKARI$ NEP wide field can provide accurate photometric redshift of the $AKARI$ sources. The 5$\sigma$ detection depths of HSC (over a 1 arcsec aperture) at $g$, $r$, $i$, $z$, and $Y$-bands are 28.6, 27.3, 26.7, 26.0, and 25.6 mag, respectively.

The u$^{\ast}$ imaging data of Canada-France-Hawaii Telescope \citep[CFHT; ][]{Iye2003} MegaCam is collected from \citet{Hwang2007} and \citet{Oi2014}.
The data of u imaging data of CFHT MegaPrime is obtained from \citet{Goto2019} and \citet{Huang2020}.

We also obtained data from Maidanak's Seoul National University 4K $\times$ 4K Camera \citep[SNUCAM; ][]{Jeon2010}. The SNUCAM contains $B$, $R$, and $I$ imaging, where the $B$, $R$, and $I$- band data reach the depths of $\sim$23.4, $\sim$23.1, and $\sim$ 22.3 mag at 5$\sigma$, respectively.

We also used far-IR data from $Herschel$/PACS \citep{Pearson2019} and $Herschel$/SPIRE (Pearson et al. in prep) observations at the NEP region. The PACS observation covers $\sim$0.44 deg$^2$ overlapping with the NEP region, which contains the 100 $\mu$m and 160 $\mu$m imaging data. SPIRE covers 9 deg$^2$ of the NEP, containing 250 $\mu$m, 350 $\mu$m and 500 $\mu$m imaging data. 
\begin{table*}
	\centering
	\caption{Summary of the 36-band data used in this work.}
	\label{tab:band}
	\begin{tabular}{lccccr} 
		\hline
		Instrument & \makecell[c]{Area \\(deg$^{2}$)} & Filter & \makecell[c]{Effective wavelength \\($\mu m$)} & \makecell[c]{Detection limit \\(5$\sigma$, AB mag)} & References\\
		\hline
		$AKARI$/IRC & 5.4 & $N2$ & 2.3 & 20.9 & \cite{Kim2012}\\
		 & 5.4 & $N3$ & 3.2 & 21.1 & \cite{Kim2012}\\
		 & 5.4 & $N4$ & 4.1 & 21.1 & \cite{Kim2012}\\
		 & 5.4 & $S7$ & 7 & 19.5 & \cite{Kim2012}\\
		 & 5.4 & $S9W$ & 9 & 19.3 & \cite{Kim2012}\\
		 & 5.4 & $S11$ & 11 & 19.0 & \cite{Kim2012}\\
		 & 5.4 & $L15$ & 15 & 18.6 & \cite{Kim2012}\\
		 & 5.4 & $L18W$ & 18 & 18.7 & \cite{Kim2012}\\
		 & 5.4 & $L24$ & 24 & 17.8 & \cite{Kim2012}\\
		\hline
		$Subaru$/HSC & 5.4 & $g$ & 0.47 & 28.6 & Oi et al. (2020, submitted)\\
		 & 5.4 & $r$ & 0.61 & 27.3 & Oi et al. (2020, submitted)\\
		 & 5.4 & $i$ & 0.76 & 26.7 & Oi et al. (2020, submitted)\\
		 & 5.4 & $z$ & 0.89 & 26.0 & Oi et al. (2020, submitted)\\
		 & 5.4 & $Y$ & 0.99 & 25.6 & Oi et al. (2020, submitted)\\
		\hline 
		CFHT/MegaCam & \makecell[c]{2, \\0.7} & $u^{*}$ & 0.39 & \makecell[c]{26, \\24.6} & \makecell[r]{\cite{Hwang2007}, \\\cite{Oi2014}}\\ 
		& 3.6 & u & 0.38 & 25.27 & \citet{Huang2020}\\
		\hline 
		CFHT/WIRCam & 0.7 & $Y$ & 1.02 & 23.4 & \cite{Oi2014}\\
		& 0.7 & $J$ & 1.25 & 23.0 & \cite{Oi2014}\\
		& 0.7 & $K_{s}$ & 2.14 & 22.7 & \cite{Oi2014}\\
		\hline
		Maidanak/SNUCAM & 4 & $B$ & 0.44 & 23.4 & \cite{Jeon2010}\\
		& 4 & $R$ & 0.61 & 23.1 & \cite{Jeon2010}\\
		& 4 & $I$ & 0.85 & 22.3 & \cite{Jeon2010}\\
		\hline
		KPNO/FLAMINGOS & 5.1 & $H$ & 1.6 & 21.3 & \cite{Jeon2014}\\
		\hline
		$Spitzer$/IRAC & 7. & IRAC1 & 3.6 & 21.8 & \cite{Nayyeri2018}\\
		& 7 & IRAC2 & 22.4 & 4.5 & \cite{Nayyeri2018}\\
		& 0.4 & IRAC3 & 5.8 & 16.6 & \cite{Jarrett2011} \\
		& 0.4 & IRAC4 & 8 & 15.4 & \cite{Jarrett2011}\\
		\hline
        $WISE$/ALLWISE & 1.5 & $W1$ & 3.4 & 18.1 & \cite{Jarrett2011}\\
		& 1.5 & $W2$ & 4.6 & 17.2 & \cite{Jarrett2011}\\
		& 1.5 & $W3$ & 12 & 18.4 & \cite{Jarrett2011}\\
		& 1.5 & $W4$ & 22 & 16.1 & \cite{Jarrett2011}\\
		\hline
		$Herschel$/PACS & 0.44 & Green & 100 & 14.7 & \citet{Pearson2019}\\
		& 0.44 & Red & 160 & 14.1 & \citet{Pearson2019}\\
		\hline
		$Herschel$/SPIRE & 9 & PSW & 250 & 14 & Pearson et al. in prep.\\
		& 9 & PMW & 350 & 14.2 & Pearson et al. in prep.\\
		& 9 & PLW & 500 & 13.8 & Pearson et al. in prep.\\
		\hline
	\end{tabular}
\end{table*}

\subsection{SED fitting using CIGALE}

We used \begin{footnotesize}CIGALE\end{footnotesize} version 2018.0. to perform SED fitting and model the optical to far-IR emission of each source. The model is based on the energy balance principle: the UV-to-optical energy absorbed by dust is self-consistently re-emitted in the mid- to far-IR.
\begin{footnotesize}CIGALE\end{footnotesize} allows many parameters to be varied in order to fit an observed SED, such as star formation history (SFH), single stellar population (SSP), attenuation law, AGN emission, and dust thermal emission.

We assumed a SFH delayed with optional exponential burst. We fixed e-folding times of the main stellar population ($\tau_{\rm main}$) and the late starburst population ($\tau_{\rm burst}$), while the age of the main stellar population in the galaxy is parameterised. We adopted the stellar templates from \citet{Bruzual2003}, by assuming the initial mass function (IMF) introduced by \citet{Salpeter1955}, and the standard default nebular emission model in \begin{footnotesize}CIGALE\end{footnotesize} \citep[see also][]{Inoue2011}.
Dust attenuation is modelled following \citet{Charlot2000} with additional flexibility. 
The \citet{Charlot2000} attenuation law recipe is described by two power laws: one for birth cloud (BC), and the other for the interstellar medium (ISM). According to the original law, both power law slopes are equal to $-0.7$. In our analysis, we introduce slopes flexibility (then the law becomes more similar to the Double Power Law, see for example \citet{LoFaro2017} and \citet{Buat2018}. We also separately parameterised the V-band attenuation in the ISM ($A_{\rm V}^{\rm ISM}$). We do not introduce any flexibility for the $\mu$ parameter which is defined as the ratio of the attenuation in the V-band experienced by old and young stars. 
While earlier works on nearby starburst galaxies obtained $\mu = 0.3$ \citep[e.g.][]{Charlot2000}, recent works found slightly higher values of $\mu = 0.44$ \citep[e.g.][]{Malek2018} and $\mu = 0.5$ \citep[e.g.][]{Buat2019}. For our analysis we set $\mu=0.44$ \citep{Malek2018}. The reprocessed IR emission of dust absorbed from UV/optical stellar emission is modelled assuming dust templates of \citet{Draine2014}.
For AGN emission, we used models provided by \citet{Fritz2006}. We fixed certain parameters that determine the number density distribution of the dust within the dust torus , i.e., ratio of the maximum to minimum radii of the torus ($R_{\rm Max}/R_{\rm min}$), density profile along the radial and the polar distance coordinates parameterised by $\beta$ and $\gamma$ (see equation 3 in \citet{Fritz2006}), and opening angle. We parameterised the optical depth at 9.7 $\mu$m ($\tau$) and $\psi$ parameter (an angle between equatorial axis and line of sight) that corresponds to our viewing angle of the torus. We also parameterise AGN fraction ($\rm f_{\rm AGN\_IR}$) as the contribution of IR luminosity from AGN to the total IR luminosity:

\begin{equation} \label{eq:1}
L _{\rm AGN}= L _{\rm TIR} \times \rm f_{\rm AGN\_IR},
\end{equation}
where $L_{\rm AGN}$ is the AGN IR luminosity, and $\rm f_{\rm AGN\_IR}$ is the contribution of the AGN to the total IR luminosity ($L_{\rm TIR}$). We did not set AGN contribution up to 1. In \begin{footnotesize}CIGALE\end{footnotesize}, it is practically very difficult to obtain a pure AGN with AGN contribution equal to 1. A source with AGN contribution equal to 0.7 is already very large, and some literature investigates AGNs with \begin{footnotesize}CIGALE\end{footnotesize} by using the same upper limit of AGN contribution as ours \citep[e.g., ][]{Eser2020}.  Therefore, we set this parameter ranging only from 0 to 0.7. Also, in order to avoid misunderstandings from other literature \citep[e.g., ][]{Chiang2019}. who defined their AGN fraction as the number fraction of AGNs over all galaxies), we describe our AGN fraction from the SED fitting as `AGN contribution' in this work.

The parameter settings we used in \begin{footnotesize}CIGALE\end{footnotesize} are summarised in Table~\ref{tab:CIGALE_parameters}.

\subsection{Sample Selection}

In order to obtain good fitting results, the mid-IR and far-IR photometry is important. We require a detection in the far-IR wavelengths, particularly in $Herschel$ SPIRE 250 $\mu$m or PACS 100 $\mu$m band. In addition to the requirement of detection in the far-IR, we also selected sources detected in $AKARI$ $L18W$ band since the detection depth in $Herschel$ $SPIRE$ and $PACS$ are too shallow and the uncertainty in the far-IR is too large for us to have a well constrained SED fitting. The $L18W$ passband is wider than the $L15$ and $L24$ bands and it has higher sensitivity than the $L15$ and $L24$ bands. Even though this criterion will make our sample sizes smaller, the performance of the continuum fitting by \begin{footnotesize}CIGALE\end{footnotesize} can be better constrained.
Also, all sources must have redshifts larger than 0 in order to exclude stars. We use the spectroscopic redshifts if the sources have them. For the sources with no spectroscopic redshifts, we use the photometric redshifts calculated by \citet{Ho2020}. \citet{Ho2020} also suggest that there are some sources with bad photometry of HSC, which we excluded in our work.
Furthermore, we also exclude the stars based on the star/galaxy separation by \citet{Ho2020}. 
After applying the above criteria, there are 446 sources with spectroscopic redshifts and 1397 sources with photometric redshifts.
Under the parameter settings described in Table~\ref{tab:CIGALE_parameters}, we fit the stellar, AGN, and SF components to at most 36 photometric points.

After the SED fitting, we applied a basic criterion to select the samples so that we could eliminate objects without enough information and ensure that the data are moderately well-fitted with a combination of the stellar, AGN, and SF components by \begin{footnotesize}CIGALE\end{footnotesize}.
The criterion is: 
the SED-fitting result of an object must have reduced $\chi^{2}$ < 10. The reason why we determined this selection criterion is because of the large number of data points we use in \begin{footnotesize}CIGALE\end{footnotesize}. Since there are at most 36 photometric points, the photometric points are very likely to be correlated. Therefore, the reduced $\chi^{2}$ may not be well estimated. After the visual inspection, we found reduced $\chi^{2}$ < 10 provides a reasonable limit. 
With the above criterion, there are 404 sources with spectroscopic redshift, and 1267 sources with photometric redshift, which we define as `spec-z sample' and `photo-z sample', respectively. We combine spec-z sample and photo-z sample into `all-z sample', having 1,671 sources, in total, for further analyses.
The reason why we have to separate spec-z sample and photo-z sample first is to check the AGN recovery rate in spec-z sample and reliability for the photo-z sample . As mentioned before, most of the sources with spectroscopic redshifts have already been classified as AGNs or SFGs. We first check our SFG-AGN diagnostics with the sources with spectroscopic redshifts (spec-z sample), and then apply the AGN selection based on our method to all selected sources (all-z sample). Fig.~\ref{fig:fig2} shows a flowchart of our sample selection process.

In order to ensure the reliability of the physical properties obtained from \begin{footnotesize}CIGALE\end{footnotesize}, we also took advantage of the mock analysis of the physical properties in \begin{footnotesize}CIGALE\end{footnotesize}, e.g., AGN contribution, SFR, etc. More information on the mock analysis is in Appendix~\ref{aa}.

\begin{table*}
    	\centering
    	\caption{
    	Modules and parameter values used to model the sample in CIGALE.
        }
    \label{tab:CIGALE_parameters}
    \begin{flushleft}
    \begin{tabular}{cl}\hline
    	Parameter & \multicolumn{1}{c}{Value}\\ \hline \relax
    	& \multicolumn{1}{c}{Delayed SFH}\\ \hline \relax
    	$\tau_{\rm main}$$^{*1}$ (10$^{6}$  years) & 5000.0 \\
		age$_{\rm main}$$^{*2}$ (10$^{6}$  years) & 1000, 5000, 10000\\
		tau$_{\rm burst}$$^{*3}$ (10$^{6}$  years) & 20000\\
		age$_{\rm burst}$$^{*4}$ (10$^{6}$  years) & 20\\
		f$_{\rm burst}$$^{*5}$ & 0.00,0.01\\
		sfr\_A$^{*6}$(M$_{sun}$/ year) & 1.0\\
		\hline \relax
		& \multicolumn{1}{c}{SSP \citep{Bruzual2003}} \\ \hline \relax
		Initial Mass Function & \citep{Salpeter1955} \\
	    metallicity & 0.02 \\
		separation\_age$^{*7}$ & 10 \\
		\hline \relax
		& \multicolumn{1}{c}{Dust attenuation \citep{Charlot2000}} \\ \hline \relax
		log $A_{\rm V}^{\rm ISM}$$^{*8}$& -2, -1.7, -1.4, -1.1, -0.8, -0.5, -0.2, 0.1, 0.4, 0.7, 1 \\
		$\mu$$^{*9}$ & 0.44 \\
		slope\_ISM$^{*10}$ & -0.9, -0.7, -0.5 \\
		slope\_BC$^{*11}$ & -1.3, -1.0, -0.7 \\
		\hline \relax
		& \multicolumn{1}{c}{Dust emission \citep{Draine2014}} \\ \hline \relax
		$q_{\rm PAH}$$^{*12}$ & 0.47, 1.77, 2.50, 5.26 \\
		Umin$^{*13}$ & 0.1, 1.0, 10, 50 \\
		$\alpha$$^{*14}$ & 1.0, 1.5, 2.0, 2.5, 3.0 \\
		$\gamma$$^{*15}$ & 1.0 \\
		\hline \relax
    	& \multicolumn{1}{c}{AGN emission \citep{Fritz2006}} \\ \hline \relax
    	$R_{\rm Max}/R_{\rm min}$$^{*16}$ & 60.0\\
    	$\tau$$^{*17}$ & 0.3, 6.0 \\
    	$\beta$$^{*18}$ & -0.5 \\
		$\gamma$$^{*18}$& 4.0 \\
		opening angle$^{*19}$ & 100.0 \\
		$\psi$$^{*20}$ & 0.001, 60.100, 89.990 \\
		AGN contribution & \makecell[l]{0.0, 0.025, 0.05, 0.075, 0.1, 0.125, 0.15, 0.175, 0.2, 0.225,0.25, 0.275, 0.3, 0.325, 0.35, 0.375, 0.4, 0.425, \\
		0.45, 0.475, 0.5, 0.525, 0.55, 0.575, 0.6, 0.625, 0.65, 0.675, 0.7} \\
		\hline
        \end{tabular}\\
        $^{*1}$: e-folding times of the main stellar population \\
        $^{*2}$: age of the main stellar population in the galaxy \\
        $^{*3}$: e-folding time of the late starburst population \\
        $^{*4}$: Age of the late burst \\
        $^{*5}$: Mass fraction of the late burst population \\
        $^{*6}$: Multiplicative factor controlling the amplitude of SFR if normalise is False\\
        $^{*7}$: Age of the separation between the young and the old star populations \\
        $^{*8}$: V -band attenuation in the ISM \\
        $^{*9}$: the ratio of the attenuation in the V-band experienced by old and young stars \\
        $^{*10}$: Power law slope of the attenuation in the ISM \\
        $^{*11}$: Power law slope of the attenuation in the birth cloud \\
        $^{*12}$: Mass fraction of PAH \\
        $^{*13}$: Minimum radiation field \\
        $^{*14}$: Power law slope dU/dM ${\propto}$ $U^{\alpha}$ \\
        $^{*15}$: Fraction illuminated from Umin to Umax \\
        $^{16}$: ratio of the maximum to minimum radii of the torus \\
        $^{*17}$: the optical depth at 9.7 $\mu$m \\
        $^{*18}$: density profile along the radial and the polar distance coordinates parameterised by $\beta$ and $\gamma$ (see equation 3 in \citet{Fritz2006})\\
        $^{*19}$: opening angle\\
        $^{*20}$: an angle between equatorial axis and line of sight
    \end{flushleft}
\end{table*}

\section{Results} \label{sec3}
\subsection{SED results of spec-z/all-z samples}
We define sources with AGN contribution larger than or equal 0.2 in our work as AGNs (hereafter SED-AGNs). The remaining sources in our sample are classified as SFGs (hereafter SED-SFGs). We compare our SED fitting results with the spectroscopically/X-ray selected AGNs from \citet{Shim2013} and \citet{Krumpe2015}. Also we compare our results with the AGN selection using another SED fitting code, \begin{footnotesize}Le Phare\end{footnotesize} (Oi et al. submitted). Detailed discussions of these comparisons are in Sec.~\ref{4.1}.

\begin{figure}
    \centering
	\includegraphics[width=\columnwidth]{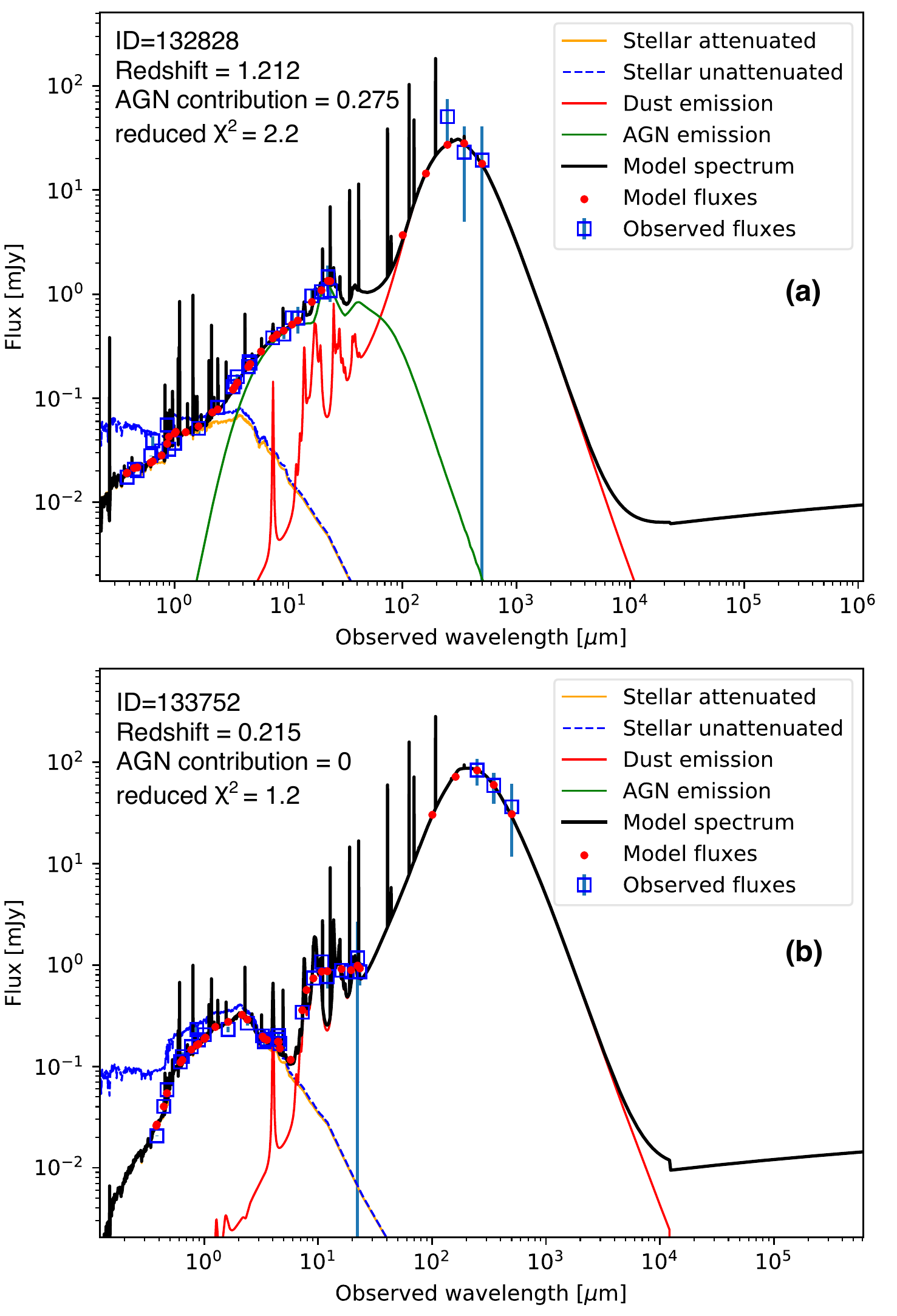}
    \caption{Examples of the SED fitting from CIGALE. (a) A typical SED of AGN/composite galaxies. (b) A typical SED of SFGs.}
    \label{fig:fig3}
\end{figure}

According to our definition of AGNs, we found 24 AGNs from the spec-z sample. Also, 102 AGNs are found in the photo-z sample. Therefore, there are 126 AGNs identified in this work.
On average, the IR SED of typical AGNs (i.e., $2-10$ keV luminosity, L$_{2 - 10\,{\rm keV}}$ $\sim$10$^{42} - 10^{44}$ erg s$^{-1}$) is best described as a broken power-law at $\leq$ 40 $\mu$m that falls steeply at far-IR wavelengths \citep{Mullaney2011}. We checked the SEDs of our sample, and our classification is consistent with the description of IR SED in \citet{Mullaney2011}. Fig.~\ref{fig:fig3} shows examples of the SED fitting with \begin{footnotesize}CIGALE\end{footnotesize}. Fig.~\ref{fig:fig3}(a) is the SED shape of an AGN in this work. Fig.~\ref{fig:fig3}(b) is the SED shape of a SFG in this work.

\subsection{Correlation between AGN contribution and other parameters}

\begin{figure*}
    \centering
	\includegraphics[width=\linewidth]{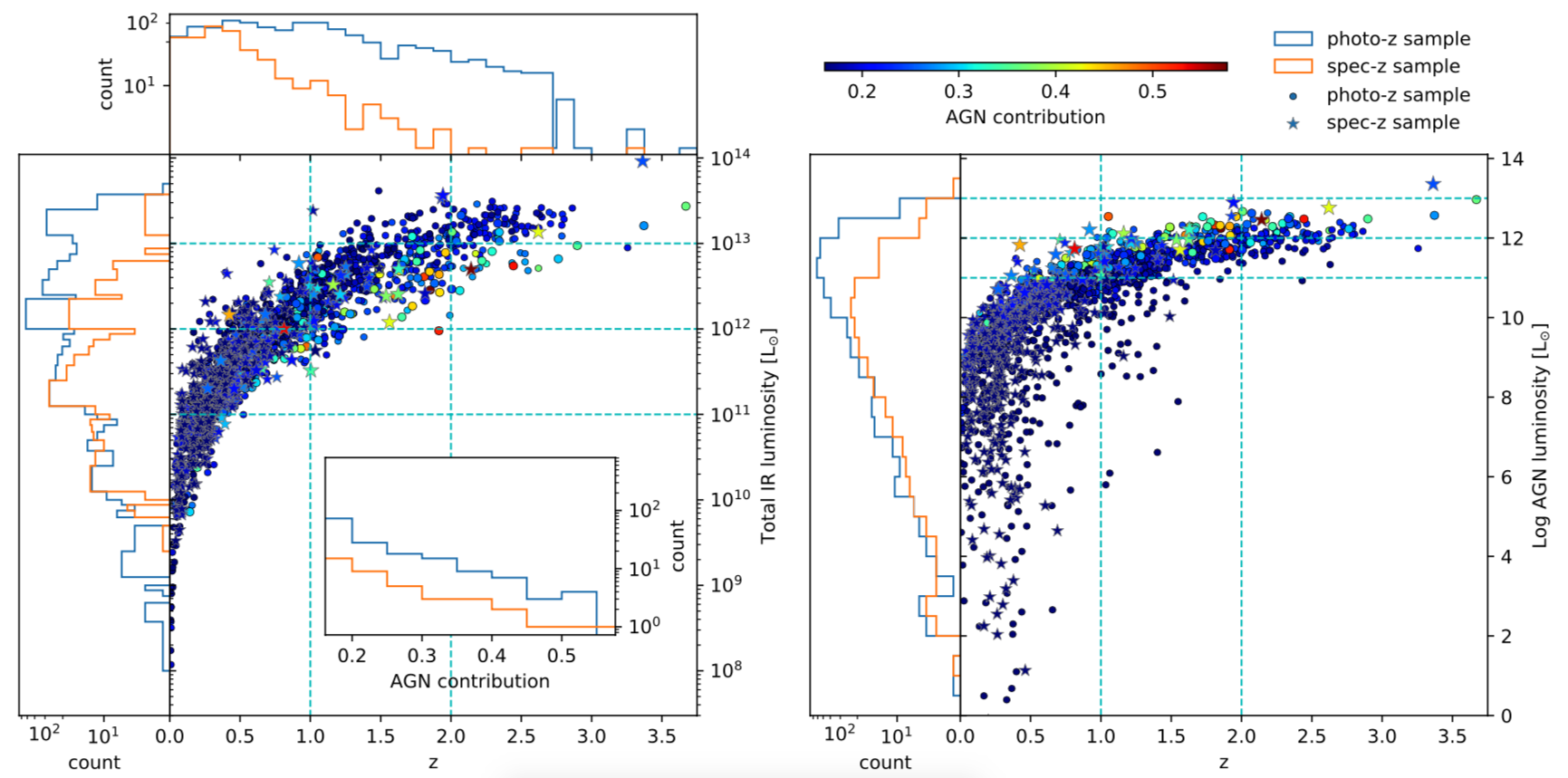}
    \caption{Luminosity against redshifts of the spec-z and photo-z samples in this work. The colours of the stars and dots represent the distribution of the AGN contribution of spec-z and photo-z samples. Left panel: Total IR uminosity against redshifts of the spec-z and photo-z samples. The histogram along with x-axis is the distribution of redshift. The histogram along with y-axis is the distribution of total IR luminosity. The inset histogram is the distribution of the AGN contribution. Right panel: AGN luminosity against redshifts of the spec-z and photo-z samples. The histogram along with y-axis is the distribution of AGN luminosity.}
    \label{fig:fig4}
\end{figure*}

One of the advantages of using \begin{footnotesize}CIGALE\end{footnotesize} is that we can obtain physical properties, e.g., AGN contribution and AGN luminosity. We obtain the values of AGN contribution and AGN luminosity, and calculate the total IR luminosity of every source by using Eq. \ref{eq:1}. We investigate whether the presence of AGNs is correlated with redshift or total IR luminosity. 
Fig.~\ref{fig:fig4} shows AGN contribution, redshift, total IR luminosity and AGN luminosity of spec-z and all-z samples, individually. It is clear that when the redshift of a source is higher, the AGN contribution is higher. Also, when the total IR luminosity and AGN luminosity of a source becomes higher, the AGN contribution is also larger. However, it is not clear whether AGN contribution depends on either redshift or total IR luminosity, or both. Therefore, in the following section, we investigate these two properties separately and see if they have a correlation with AGN contribution.

\subsubsection{AGN contribution as a function of IR luminosity}
\begin{figure}
    \centering
	\includegraphics[width=\linewidth]{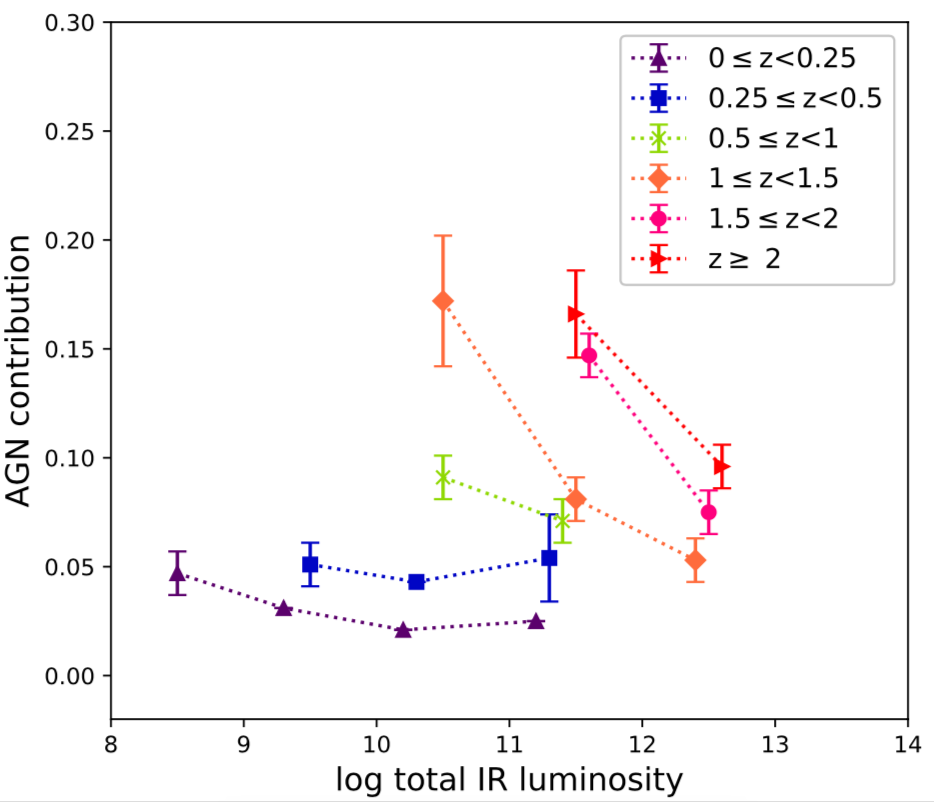}
    \caption{AGN contribution as a function of total IR luminosity of the all-z sample. The average ($\pm{1}$ standard error) AGN contributions in each bin are connected by the dashed lines. The purple triangles are the 0 $\leq$ z< 0.25 bin. The blue squares are the 0.25 $\leq$ z< 0.5 bin. Green crosses are the 0.5 $\leq$ z< 1 bin. Orange diamonds are the 1 $\leq$ z< 1.5 bin. Pink circles are the 1.5 $\leq$ z< 2 bin. Red triangles are the z$\geq$ 2 bin. }
    \label{fig:fig5}
\end{figure}

Fig.~\ref{fig:fig5} shows the AGN contribution as a function of total IR luminosity in different redshift bins of the all-z sample.
An average value of AGN contribution in each redshift bin is shown. We separate our sample by redshift in order to see if AGN contribution has luminosity dependence. Fig.~\ref{fig:fig5} shows that the AGN contribution does not change much with the total IR luminosity at redshift between 0 and 0.5. The AGN contribution is lower when the total IR luminosity is higher at redshift $\geq$ 0.5. The reason why the trend of AGN contribution changes at redshift $\geq$ 0.5 may be the different sample size in different redshift bin. Also, this trend may be more apparent because the IR luminosity detection limit changes with redshift within the bin. Fig.~\ref{fig:fig5} also shows that the overall AGN contribution is higher at higher redshift bins. This result may imply that AGN contribution is higher at higher redshift at a fixed IR luminosity. 

Also, since a $Herschel$ PACS 100 $\mu$m or SPIRE 250 $\mu$m detection is required to be in the sample, our sample is flux-limited in the far-IR at redshift higher than 1. Therefore, only very luminous galaxies at high-z are selected into our sample. We caution that, at high redshift, AGN dominated objects without far-IR detection might be dropped from our sample due to the flux-limit.

\subsubsection{AGN contribution as a function of redshift}
\begin{figure}
    \centering
	\includegraphics[width=\linewidth]{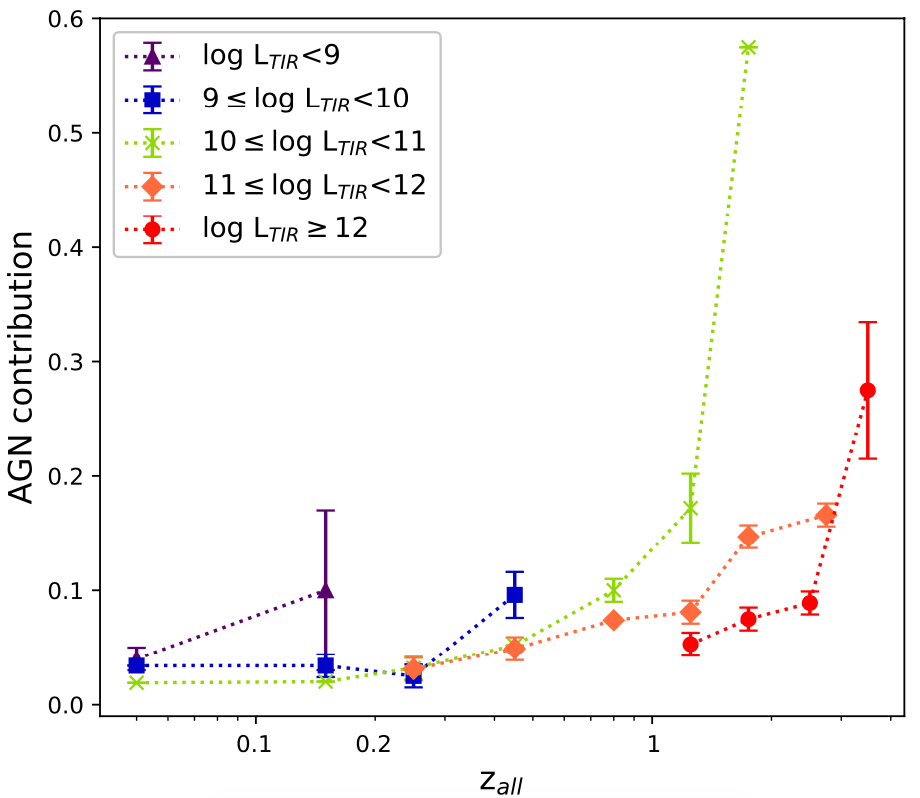}
    \caption{AGN contribution as a function of redshift of the all-z sample. The average ($\pm{1}$ standard error) AGN contributions in each bin are connected by the dashed lines. Purple triangles are the log $L_{\rm TIR}$ <9 bin. The blue squares are the 9 $\leq$ log $L_{\rm TIR}$ <10 bin. Green crosses are the 10 $\leq$ log $L_{\rm TIR}$ <11 bin. Orange diamonds are the 11 $\leq$ log $L_{\rm TIR}$ <12 bin. Red circles are the log $L_{\rm TIR}$ $\geq$ 12 bin.}
    \label{fig:fig6}
\end{figure}

Fig.~\ref{fig:fig6} shows the AGN contribution as a function of redshift with different IR luminosity bins of the all-z sample. We found that the redshift dependence is clear. When the redshift is higher, the AGN contribution becomes higher at a fixed total IR luminosity. 

Some literature discuss the relation between AGN number fraction and redshift \citep[e.g., ][]{Chiang2019}. 
The AGN number fraction is defined as:
\begin{equation} \label{eq:2}
{\rm f}_{\rm AGN\_num}= \frac{N_{\rm AGN}}{N_{\rm SFG+AGN}} ,
\end{equation}
where $\rm f_{\rm AGN\_num}$ is the AGN number fraction, $N_{\rm AGN}$ is the number of AGNs. $N_{\rm SFG+AGN}$ is the total number of SFGs and AGNs.
However, the AGN number fraction and the AGN contribution are different physical quantities. For example, it is possible that AGN number fraction is constant across the redshift, (for example $\sim$20\%) but at the same time the AGN contribution is changing across the redshift, i.e., still only 20\% are AGN in number, but among these AGN, the AGN contribution is higher. The converse might be true instead: the AGN contribution is constant across the redshift, but the AGN number fraction increases, for example, the AGN contribution is all 20\%, but the number of galaxies with the AGN contribution = 20\% increases. Therefore, we believe these two are different physical quantities, and thus, we would like to measure separately, compare, and draw conclusions. Since \citet{Chiang2019} discussed the AGN number fraction in NEP Wide field, we compare our results using the AGN number fraction as well. Detailed comparison is in Section~\ref{sec4.3.}.

\section{Discussion} \label{sec4}

\subsection{Reliability of CIGALE AGN Selection} \label{4.1}
In this section, we discuss our AGN definition using \begin{footnotesize}CIGALE\end{footnotesize}, i.e., AGN contribution higher than 0.2, in comparison to spectroscopic and X-ray definitions of AGNs.

Most of the sources in the spec-z sample are already classified by the spectroscopic and Chandra X-ray surveys in \citet{Shim2013} and \citet{Krumpe2015}. Some X-ray classified sources which are outside of the X-ray observation region covered by \citet{Krumpe2015} are obtained from $ROSAT$ all sky survey \citep{Gioia2003}. Sources detected in X-rays with an X-ray luminosity of log L$_{X}$>41.5 erg s$^{-1}$ (0.5-7 keV) are classified as AGNs. Also, from the spectroscopic surveys, sources with at least one emission line of FWHM>1000 km s$^{-1}$, or if they are Seyfert 1.5 type, they are selected as AGNs. The spectroscopically or X-ray selected AGNs are described as `XS-AGNs' in this regard.

\begin{figure}
    \centering
	\includegraphics[width=\linewidth]{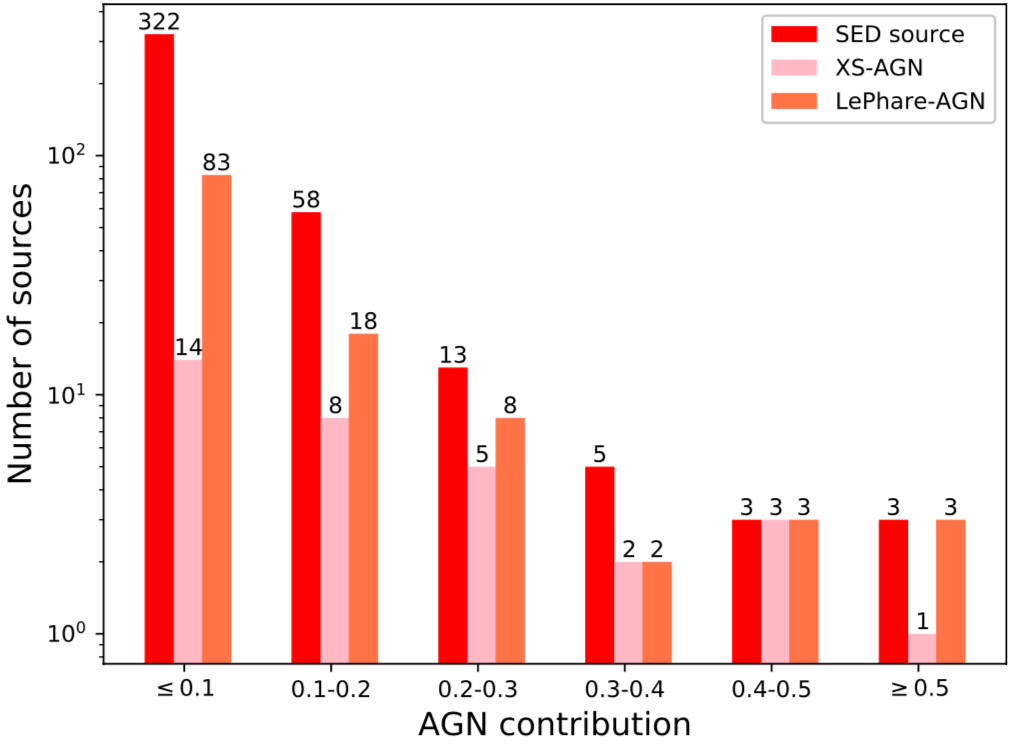}
    \caption{Distribution of the sources at different AGN contribution bin from the spec-z sample. Along x axis, there are 6 bins of different AGN contribution. Red bars are the number of the sources in different AGN contribution bins with our method. Pink bars are the number of XS-AGNs in different AGN contribution bins. Orange bars are the number of LePhare-AGNs in different AGN contribution bins.}
    \label{fig:fig7}
\end{figure}

Oi et al (submitted) also selected AGNs in the NEP-Wide field by using the observational data in band-merged catalogue from Kim et al. (submitted), and classified AGNs by using the $Spitzer$ Wide-area Infrared Extragalatic (SWIRE) templates from \citet{Polletta2007} in \begin{footnotesize}Le Phare\end{footnotesize}, which includes 16 galaxy and starburst templates and 9 AGN templates. In this regard, we describe the classification as `LePhare-AGN'.
We compare our results for the spec-z sample with the previous classifications. Fig.~\ref{fig:fig7} shows number of sources in different AGN contribution bins of the spec-z sample. 
\citet{Shen2020} selected AGNs by using \begin{footnotesize}CIGALE\end{footnotesize}. They investigated 179 radio-IR galaxies drawn from a sample of spectroscopically confirmed galaxies, which are detected in radio and mid-IR in the redshift range of 0.55 $\leq$ z $\leq$ 1.30 in the Observations of Redshift Evolution in Large Scale Environments (ORELSE) survey. They define AGNs as AGN contribution larger than 0.1. If we apply the definition of AGN from \citet{Shen2020} to our all-z samples, there are 82 sources meeting this AGN selection criterion. We divided the distribution of AGN contribution of the spec-z sample into different bins and compare with other AGN classifications (see Fig.~\ref{fig:fig7}, 14\% and 31\% of the sources with AGN contribution between 0.1 and 0.2 are classified as XS-AGNs and LePhare-AGNs, respectively. Therefore, in this work, we adopt a stricter definition of AGN, which is a source with AGN contribution larger than 0.2.

Due to the limited survey depth from \citet{Krumpe2015}, some AGNs may be missed in X-ray sources. In the case of spectroscopically selected AGNs, these are not complete since it is not possible to place a fiber on every IR-selected source due to the limited number of fibers.

In this work, we made a more strict definition of SED-AGNs. We define our SED-AGNs as sources with AGN contribution $\geq$ 0.2. The reason why sources with AGN contribution between 0.1 and 0.2 are not defined as SED-AGNs is because 69\% percent of them are not classifed as XS-AGNs or LePhare-AGNs. Therefore, only sources with AGN contribution $\geq$ 0.2 are selected as SED-AGNs. There are 24 and 102 SED-AGNs that we found in the spec-z and photo-z samples, respectively. In total, 126 SED-AGNs are found in our all-z sample.

\subsection{Comparison between colour-selected AGNs and `SED-AGNs'}

\begin{figure}
    \centering
	\includegraphics[width=\linewidth]{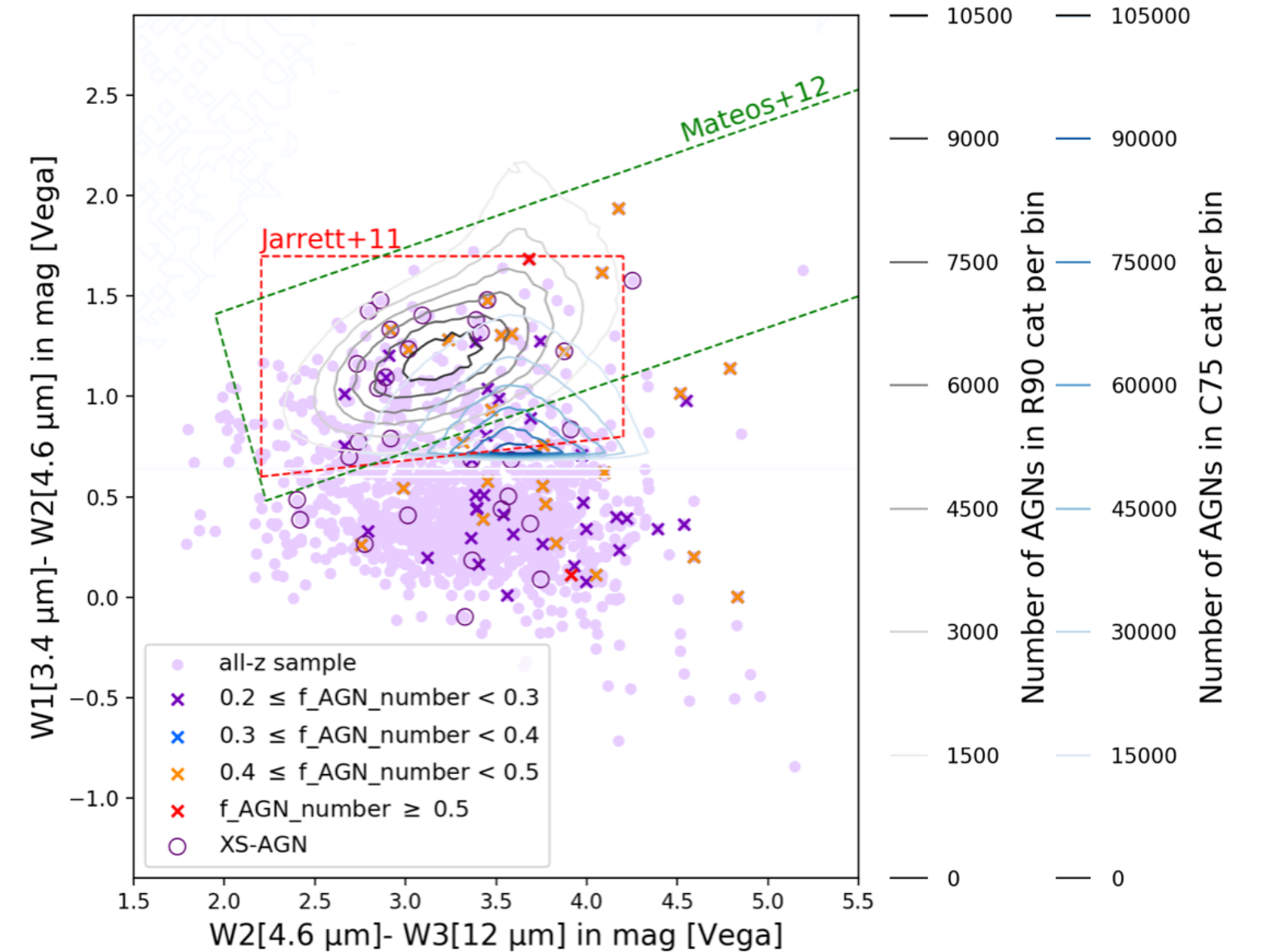}
    \caption{The $WISE$ colour-colour diagram of the all-z sample. The red-lined box is the AGN criteria from \citet{Jarrett2011}. The green-lined box is the AGN criteria from \citet{Mateos12}. The light purple dots are the all-z sample. The black circles are XS-AGNs. The grey and blue contours are the distribution of AGNs in the R90 and C75 catalogues from \citet{Assef2018}. The dark purple, blue, orange and red crosses are 0.2$\leq$ f$_{\rm AGN\_IR}$ < 0.3, 0.3$\leq$ f$_{\rm AGN\_IR}$ < 0.4, 0.4$\leq$ f$_{\rm AGN\_IR}$ < 0.5 and f$_{\rm AGN\_IR}$ $\geq$ 0.5, respectively.}
    \label{fig:fig8}
\end{figure}

We compare our results with the AGN classifications defined by \citet{Jarrett2011} and \citet{Mateos12}. \citet{Jarrett2011} and \citet{Mateos12} used $WISE$ colours to classify AGNs. In Fig.~\ref{fig:fig8}, the red box is the `AGN' region defined in \citet{Jarrett2011}, encompassing quasars (QSOs) and Seyfert galaxies. However, inside the $WISE$ colour box in both figures, there are still many SED-AGNs. Furthermore, many XS-AGNs are outside the $WISE$ colour box. 
Recently, \citet{Assef2018} released catalogues of two AGN candidates, R90 catalogue and C75 catalogue, from the $WISE$ ALLWISE Data Release based on the W1 and W2 colour-magnitude selection. Both catalogues are selected purely using the $WISE$ W1 and W2 bands (see Equation 4 and 5 in \citet{Assef2018}). The R90 catalogue consists of 4,543,530 AGN candidates with 90\% reliability, while the C75 catalogue consists of 20,907,127 AGN candidates with 75\% completeness. These reliability and completeness figures were determined from a detailed analysis of UV- to near-IR spectral energy distributions of ~105 sources in the 9 deg$^{2}$ Bo{\"o}tes field. To examine whether the AGN classifications defined by \citet{Jarrett2011} and \citet{Mateos12} match the two AGN catalogues from \citet{Assef2018}, we overplot the two AGN catalogues and compare with our SED fitting result in Fig.~\ref{fig:fig8}. The grey and blue contours in Fig.~\ref{fig:fig8} are the distribution of $WISE$ R90 and $WISE$ C75 AGN catalogues. Even R90 catalogue is with 90\% reliability and C75 catalogue is with 75\% completeness, not all the AGNs are selected by the two AGN boxes. Also, not all the SED-AGNs in our all-z sample are selected by the two AGN boxes. This shows the limitation in selecting AGNs only using two bands, W1 and W2. In our work, we use at most 36 bands, including 17 in mid-IR, allowing us more sophisticated AGN selection through the advanced SED fitting.

\begin{figure}
    \centering
	\includegraphics[width=\linewidth]{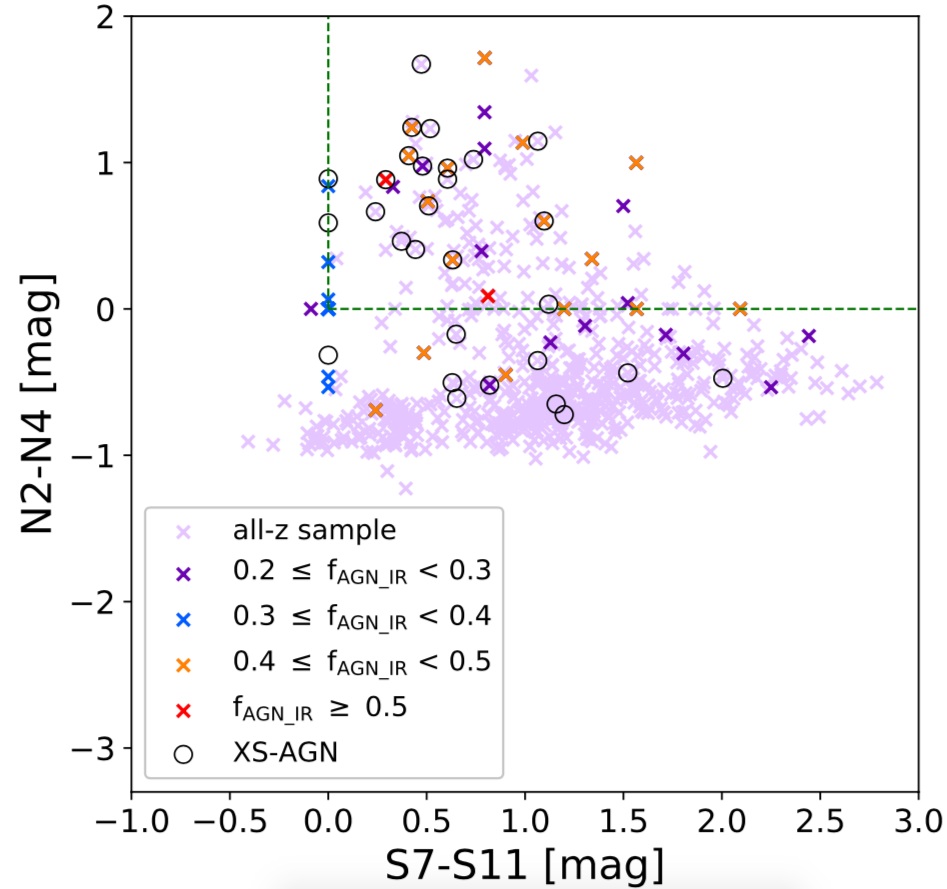}
    \caption{The $AKARI$ colour-colour diagram of the all-z sample. The upper right green-lined box is the AGN selected from \citet{Lee2007}. The light purple dots are the all-z sample. The dark purple circles are XS-AGNs. The dark purple, blue, orange and red crosses are 0.2$\leq$ f$_{\rm AGN\_IR}$ < 0.3, 0.3$\leq$ f$_{\rm AGN\_IR}$ < 0.4, 0.4$\leq$ f$_{\rm AGN\_IR}$ < 0.5 and f$_{\rm AGN\_IR}$ $\geq$ 0.5, respectively.}
    \label{fig:fig9}
\end{figure}

\citet{Lee2007} used $AKARI$ colours to classify AGNs. Sources with $AKARI$ $S7$-$S11$> 0 mag and $N2$-$N4$> 0 mag are defined as AGNs. In Fig.~\ref{fig:fig9}, we compare our results with \citet{Lee2007}. The red-lined box is the AGN criteria from \citet{Lee2007}. The black circles are the XS-AGNs. In terms of $AKARI$ $S7$-$S11$ colour the AGN criteria box is mostly consistent with our SED AGN selection. However, in terms of $AKARI$ $N2$-$N4$ colour, there are still many AGNs outside of the colour range.   

The reason why there some AGNs outside the region satisfying the AGN selection criteria \citep{Jarrett2011, Mateos12, Lee2007} is because those authors proposed the criteria on the basis of a compromise between the completeness and the reliability of AGN selection. If simple completeness (i.e., to identify more AGNs) is increased, the reliability of the selection criteria would decreased because of contamination.

\begin{figure}
    \centering
	\includegraphics[width=\columnwidth]{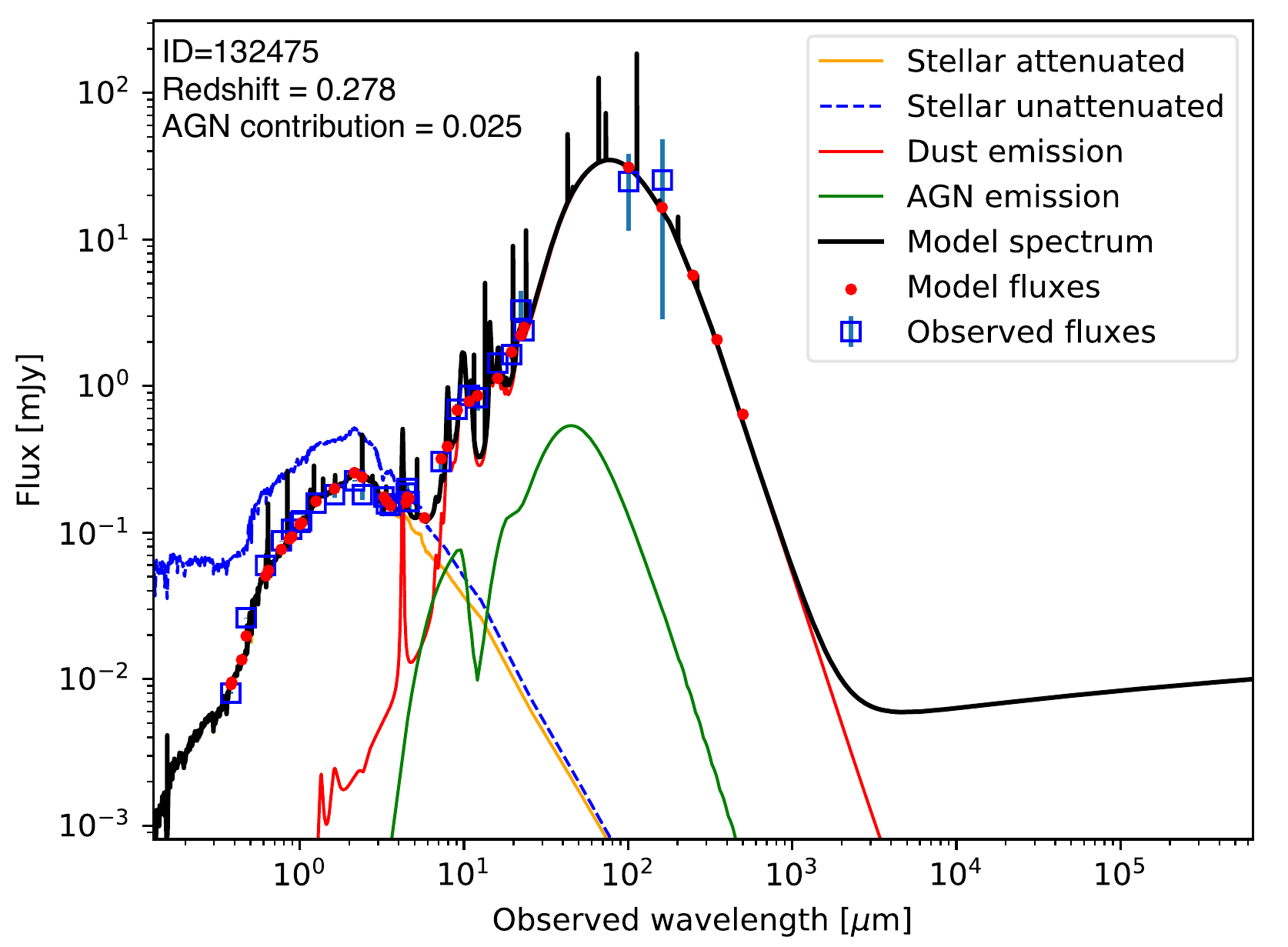}
    \caption{An example SED fitting result from CIGALE. ID = 132475 is more likely to be an SFG than an AGN.}
    \label{fig:fig10}
\end{figure}

\begin{figure}
    \centering
	\includegraphics[width=\linewidth]{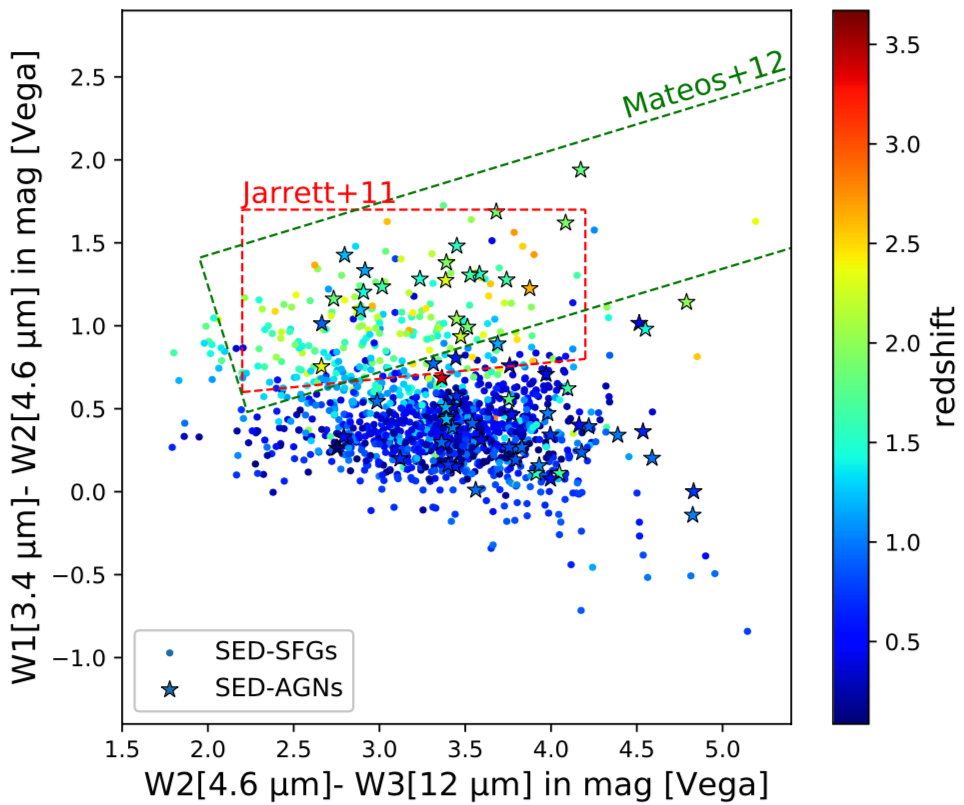}
    \caption{The WISE colour-colour diagram of the all-z sample. The red-lined box is the AGN criteria from \citet{Jarrett2011}. The green-lined box is the AGN criteria from \citet{Mateos12}. The coloured dots are the SED-SFGs. The coloured stars are the SED-AGNs. The different colours of stars and dots represent different redshifts.}
    \label{fig:fig11}
\end{figure}

Also, there are many SED-SFGs inside the $WISE$ colour box and $AKARI$ colour boxes. We checked the SEDs of some SED-SFGs which are inside all of the colour-colour boxes. These objects do not have obvious AGN power law features but have PAH emission feature peaks. Therefore, these sources are more likely to be SFGs (see Fig.~\ref{fig:fig10}). This may imply that we may encounter more difficulties when we identify SFGs and AGNs using colour-colour diagrams. Since colours are often affected by redshifts, we plot the $WISE$ colour-colour diagram with redshift. Fig.~\ref{fig:fig11} shows the redshift distribution of the sources in the $WISE$ colour-colour diagram. There are many high-z SED-SFGs inside the AGN criteria boxes. On the other hand, there are many low-z SED-AGNs outside the AGN criteria boxes. This result may imply that we may fail to detect low-z AGNs if we use colour-colour diagrams.

\subsection{AGN number fraction} \label{sec4.3.}

\begin{figure}
    \centering
	\includegraphics[width=\linewidth]{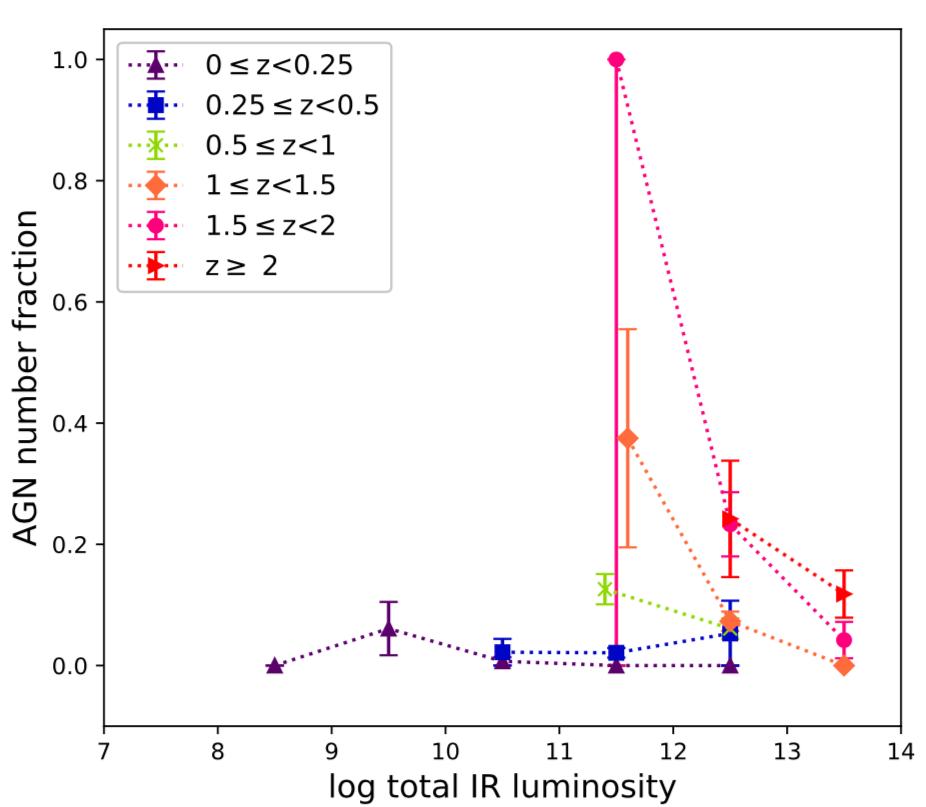}
    \caption{AGN number fraction as a function of total IR luminosity in the all-z sample. Purple triangles are the 0 $\leq$ z < 0.25 bin. Blue squares are the 0.25 $\leq$ z < 0.5 bin. Green crosses are the 0.5 $\leq$ z < 1 bin. Orange diamonds are the 1 $\leq$ z < 1.5 bin. Pink circles are the 1.5 $\leq$ z < 2 bin. Red circles are the z $\geq$ 2 bin.}
    \label{fig:fig12}
\end{figure}

\begin{figure}
    \centering
	\includegraphics[width=\linewidth]{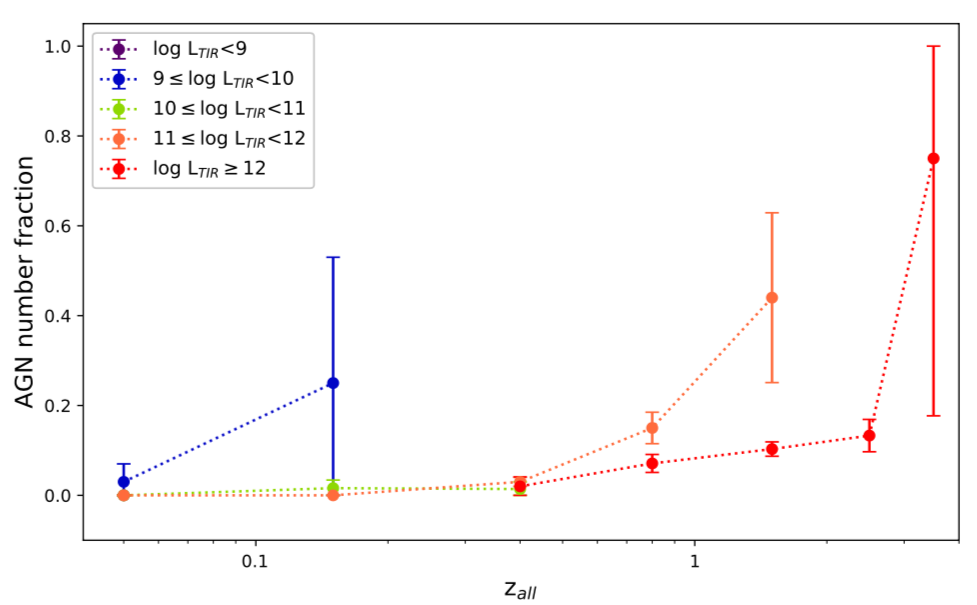}
    \caption{AGN number fraction as a function of redshift in the all-z sample. The purple triangles are the log $L_{\rm TIR}$ < 9 bin. The blue squares are the 9 $\leq$ log $L${\tiny TIR} < 10 bin. Green crosses are the 10 $\leq$ log $L_{\rm TIR}$ < 11 bin. Orange diamonds are the 11 $\leq$ log $L_{\rm TIR}$ < 12 bin. Red circles are the log $L_{\rm TIR}$ $\geq$ 12 bin.}
    \label{fig:fig13}
\end{figure}

To test theories of galaxy-SMBH co-evolution and to detect the sources of the observed IR and hard X-ray backgrounds, the AGN number fraction is crucial \citep{Rovilos2014, Shi2013a}. We plot our SED-AGN number fraction as a function of total IR luminosity and redshift and compare our results with \citet{Chiang2019}.
\citet{Chiang2019} used \begin{footnotesize}Le Phare\end{footnotesize} to select AGNs in the NEP Wide field with 18 IR bands of data, including $AKARI$'s 9 passbands, $WISE$ 1-4, $Spitzer$ IRAC 1-4 and MIPS 1 photometry. Their results indicate that AGN number fraction seems to show stronger IR luminosity dependence than redshift dependence. They also examined the fractions of SFGs and found mild decreasing trends at high IR luminosities.
Fig.~\ref{fig:fig12} shows AGN number fraction as a function of total IR luminosity. There is no clear relation between AGN number fraction and total IR luminosity at a fixed redshift.
Fig.~\ref{fig:fig13} shows AGN number fraction as a function of redshift. With the increasing redshift, the AGN number fraction becomes higher at a fixed luminosity.

Our results show the opposite conclusion from \citet{Chiang2019}. We attempt to propose some possibilities regarding this discrepancy.
We argue the differences between our results and \citet{Chiang2019} are mainly due to different sample selection and different SED-fitting techniques. Even though both samples are in the NEP wide field, we only select $AKARI$ L18W and $Herschel$ PACS 100$\mu$m or SPIRE detected sources, whereas \citet{Chiang2019} selected all of the $AKARI$ sources regardless of L18W detection. Furthermore, \citet{Chiang2019} only included $AKARI$ 9 passbands, $WISE$ 1-4, $Spitzer$ IRAC 1-4 and MIPS 1 photometry. They did not have any far-IR photometry to construct a better constraint on SED fitting. All the sources in our sample are detected in optical, mid-IR and far-IR, which allow us to avoid having low-quality SED fitting. We also use different SED-fitting techniques. \citet{Chiang2019} used \begin{footnotesize}Le Phare\end{footnotesize} to select AGNs. They used the \citet{Polletta2007} templates, which only contains 3 elliptical, 7 spiral, 6 starburst, 7 AGN, and 2 AGN- Starburst composite models, to determine the spectral type of galaxies. Meanwhile, we use \begin{footnotesize}CIGALE\end{footnotesize}, which enables us to construct millions of models and derive the SFG-AGN decomposition directly. Therefore, we argue different sample selection and different SED-fitting techniques cause the different results.

\subsection{Extremely luminous sources} \label{4.2}

\begin{figure*}
    \centering
	\includegraphics[width=\linewidth]{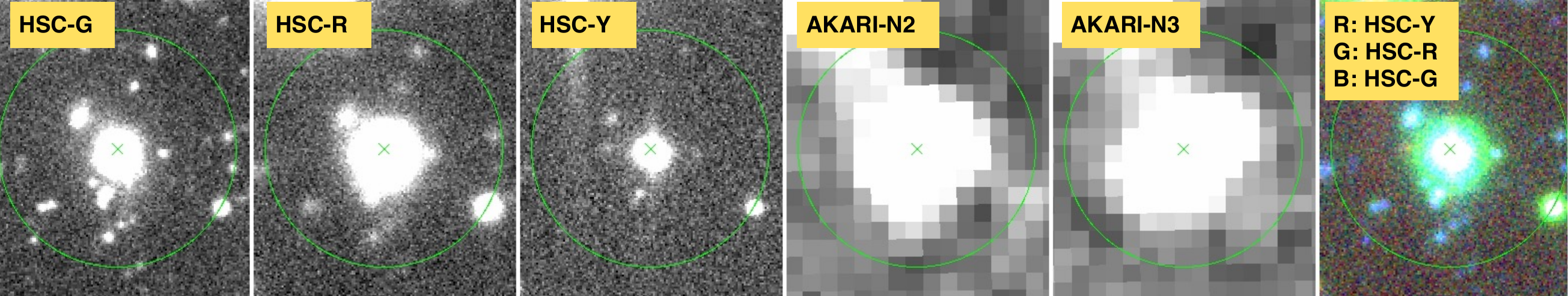}
    \caption{Images of ID=133652. The redshift of the source is 3.363, and its AGN contribution is 0.25 $\pm$ 0.06. The SFR is 1.0$^{+0.4}_{-0.4}$ $\times$ $10^{4}$ $M_{\odot} {\rm yr}^{-1}$. The total IR luminosity is 9.03 $^{+3.0}_{-1.8}$ $\times$ $10^{13}$ L$_{\odot}$. The centre of the green circle is the AKARI's coordinate. The radius of a green circle of 133652 is 10".
    }
    \label{fig:fig14}
\end{figure*}

\begin{figure*}
    \centering
	\includegraphics[width=\linewidth]{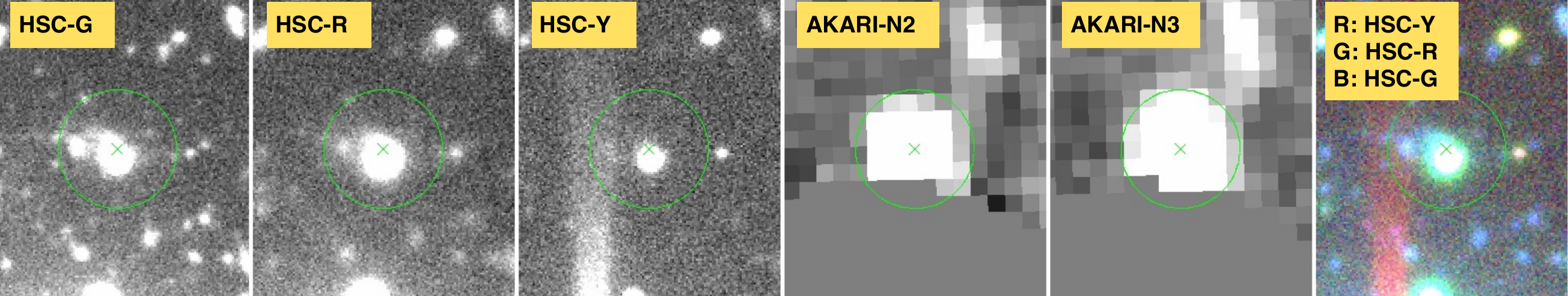}
    \caption{Images of ID=134015. The redshift of the source is 1.943, and its AGN contribution is 0.21 $\pm$ 0.02. The SFR is 2.6$^{+0.2}_{-0.2}$ $\times$ $10^{3}$ $M_{\odot} {\rm yr}^{-1}$. The total IR luminosity is 2.48 $^{+2.3}_{-1.0}$ $\times$ $10^{13}$ L$_{\odot}$. The centre of the green circle is the AKARI's coordinate. The radius of a green circle of 134015 is 5".}
    \label{fig:fig15}
\end{figure*}

\begin{figure*}
    \centering
	\includegraphics[width=\linewidth]{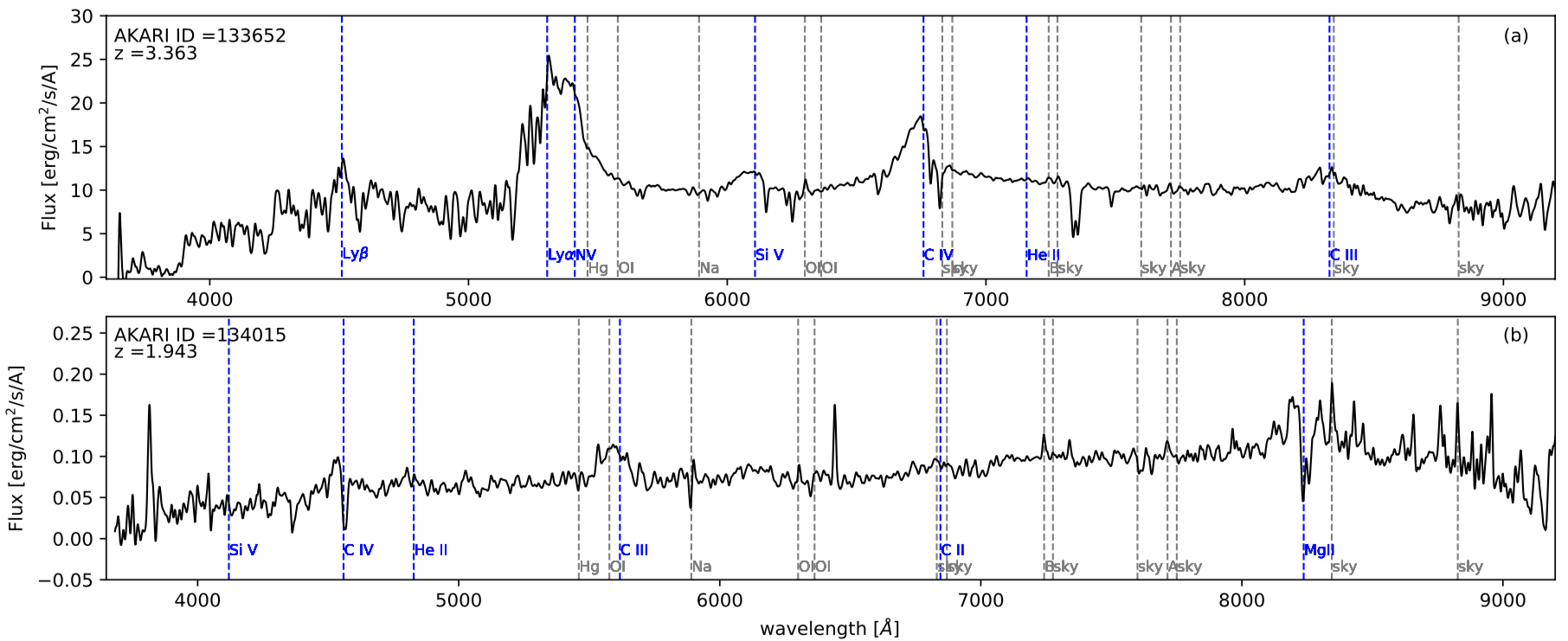}
    \caption{Spectra of 2 QSOs identified by SED fitting in this work. (a)Spectrum of ID=133652. (b)Spectrum of ID=134015. }
    \label{fig:fig16}
\end{figure*}

Galaxies with IR luminosity exceeding 10$^{13}$ L$_{\odot}$ are termed hyper luminous IR galaxies \citep[HyLIRGs;][]{Rowan-Robinson2000}. HyLIRGs are very rare populations in the local universe. However their relative abundance in galaxies is much higher at higher redshift. In addition, those luminous galaxies often show significantly high SF and/or AGN activity \citep[e.g., ][]{Fan2016, Toba2018, Wu2018, Toba2020b}, which would suggest that they correspond to the maximum phase in the growth history of galaxies. Therefore, it is important to search for IR luminous galaxies such as HyLIRGs in order to better understand the galaxy formation and evolution and connection to their SMBHs. There are 147 HyLIRGs that we found in our all-z sample. Among these HyLIRGs, we found two SED-AGNs with high SFR (> 2.5 $\times$ $10^{3}$ $M_{\odot} {\rm yr}^{-1}$). 
We checked the optical and IR images of the sources. Fig.~\ref{fig:fig14} and Fig.~\ref{fig:fig15} shows the Subaru HSC $i$, $z$, $Y$ and $AKARI$ $N2$, $N3$ images. Their spectra is in Fig.~\ref{fig:fig16}. Both sources show very strong emission in both optical and IR wavelengths. We report that these two objects are QSOs since they are point sources of high luminosity and redshift. Both sources are spectroscopically classified as Type-1 AGNs \citep{Shim2013}. With our SED-fitting method, we are able to find QSOs. The physical properties of the extremely luminous sources are summarised in Table.~\ref{tab:luminousobjects}.

We also checked the V-band attenuation in the ISM, $A_{\rm V}^{\rm ISM}$ of the HyLIRGs. We found that the average $A_{\rm V}^{\rm ISM}$ output by \begin{footnotesize}CIGALE\end{footnotesize} for objects with ID=133652 and ID=134015 are about 0.32 and 1.26, respectively. This indicates that the HyLIRG with ID=134015 is more affected by dust  extinction, and thus it may be a highly obscured HyLIRG at z = 1.9.
But, if the object with ID=134015 is affected by strong lensing, its IR luminosity should be boosted. In general, we need a high spatial resolution (< 0.1") image to justify the claim of a lensing effect. Therefore, we still need to consider other possibilities to explain why this object has a large IR luminosity.
Furthermore, the SFR of the HyLIRG with ID=133652 we obtained from \begin{footnotesize}CIGALE \end{footnotesize} is very high. We checked the relation between the IR luminosity and SFR from \citet{Kennicutt1998b}:
\begin{equation}
\label{eq:3}
{\rm SFR_{Kennicutt}} (M_{\odot} {\rm yr}^{-1})=4.5\times10^{-44}L_{\rm IR}({\rm ergs}^{-1}).
\end{equation}

The IR luminosity in \citet{Kennicutt1998b} is defined as the luminosity integrated from 8 to 1000 $\mu$m. We define the IR luminosity as dust luminosity from \citet{Draine2014} in \begin{footnotesize}CIGALE \end{footnotesize}, which is similar to $L_{IR}$ defined in \citet{Kennicutt1998b}. Then we derive the SFR from Eq.~\ref{eq:3} and check if it is consistent with the SFR we obtained from \begin{footnotesize}CIGALE \end{footnotesize}. The SFR of ID=133652 derived from \citet{Kennicutt1998b} is 7.1$^{+1.4}_{-2.5}$ $\times$ $10^{3}$ $M_{\odot} {\rm yr}^{-1}$. The SFR of ID=133652 obtained from \begin{footnotesize}CIGALE \end{footnotesize} is 1.0$^{+0.4}_{-0.4}$ $\times$ $10^{4}$ $M_{\odot} {\rm yr}^{-1}$. The difference of SFR from \citet{Kennicutt1998b} and \begin{footnotesize}CIGALE \end{footnotesize} is $\sim$ 30 percent, which seems to be large. However, the observational uncertainty of $Herschel$/SPIRE PSW (250$\mu$m) of ID= 133652 is also large ($\sim$ 19\%). Therefore, if we take it into account, it is reasonable for the ID=133652 to have such a large SFR from \begin{footnotesize}CIGALE\end{footnotesize}.

\begin{table}
	\centering
	\caption{Physical properties of the extremely luminous sources.}
	\label{tab:luminousobjects}
	\begin{tabular}{lllll}
		\hline
		ID & z & f$_{\rm AGN\_IR}$ & $L_{\rm TIR}$ (L$_{\odot}$) & SFR ($M_{\odot} {\rm yr}^{-1}$)\\
		\hline
		133652 & 3.363 & 0.25 $\pm$0.06 & 9.03 $^{+3.0}_{-1.8}$ $\times$$10^{13}$ & 1.0$^{+0.4}_{-0.4}$ $\times$$10^{4}$ \\
		134015 & 1.943 & 0.21 $\pm$0.02 & 2.48 $^{+2.3}_{-1.0}$ $\times$$10^{13}$ & 2.6$^{+0.2}_{-0.2}$ $\times$$10^{3}$ \\
		\hline
	\end{tabular}
\end{table}

\section{Conclusions} \label{sec5}
We used the unique $AKARI$ NEP-Wide sample with 36 photometric bands to perform \begin{footnotesize}CIGALE\end{footnotesize} SED fitting and select AGNs.
We define the AGN as sources with AGN contribution larger than 0.2 in this work. With this definition, we found 126 AGNs. We also found that traditional colour-colour selection criteria could miss low-z AGNS. Furthermore, We investigate the HyLIRGS in our sample and found, at least, 2 QSOs.
We extended the SED analysis of $AKARI$ sources to the entire NEP-Wide field. Based on the SED fitting results performed with \begin{footnotesize}CIGALE\end{footnotesize}, we found that AGN contribution may be more dependent on redshift than total IR luminosity. However, we do not find any clear correlation between AGN contribution and total IR luminosity in contrast to \citet{Chiang2019}. This is probably due to the different sample selection criteria, and a more sophisticated SED fitting method used in this work. More high-z sources are needed in order to confirm whether AGN contribution still depends on redshift at a fixed IR luminosity. Therefore, we expect future deep IR surveys, i.e. James Webb Space Telescope, can help us reveal this relation.

\section*{Acknowledgements}
We are very grateful to the anonymous referee for many insightful comments. This research is based on observations with $AKARI$, a JAXA project with the participation of ESA. This work used high-performance computing facilities operated by the Centre for Informatics and Computation in Astronomy (CICA) at National Tsing Hua University (NTHU) through a grant from the Ministry of Education (MOE) of the Republic of China (Taiwan). This research was conducted under the agreement on scientific cooperation between the Polish Academy of Sciences and the Ministry of Science and Technology (MOST) of Taiwan through grant 109-2927-I-007-505. TG acknowledges the support by the MOST of Taiwan through grant 108-2628-M-007 -004 -MY3. TH and AYLO are supported by the CICA at NTHU. 
AYLO's visit to NTHU is also supported by the MOST of the Republic of China (Taiwan) grant 105-2119-M-007-028-MY3, kindly hosted by Prof Albert Kong. This equipment was funded by the MOE, the MOST of Taiwan, and NTHU. TM is supported by UNAM-DGAPA PASPA and PAPIIT IN111319 as well as CONACyT 252531. KM and AP have been supported by the National Science Centre (grants UMO-2018/30/E/ST9/00082 and UMO-2018/30/M/ST9/00757).

\section*{Data Availability}
The data is available upon request.


\bibliographystyle{mnras}
\bibliography{AGN_selection} 




\appendix
\section{Mock analysis in CIGALE} \label{aa}

One feature of \begin{footnotesize}CIGALE\end{footnotesize} is the possibility to assess whether or not physical properties can actually be estimated in a reliable way through the analysis of a mock catalogue. The idea is to compare the physical properties of the mock catalogue, which are known exactly, to the estimates from the analysis of the likelihood distribution. In \begin{footnotesize}CIGALE\end{footnotesize}, best fitting is considered for each object and an artificial catalogue is built based on the best fitting. Then, noise is injected into the fluxes of this new catalogue to simulate new observations.
Each quantity is modified by adding a value taken from a Gaussian distribution with the same standard deviation as the uncertainty on the observation. This mock catalogue is then analysed in the exact same way as the original observations. Physical properties for which the exact and estimated values are similar can be estimated reliably.

\begin{figure}
    \centering
	\includegraphics[width=\linewidth]{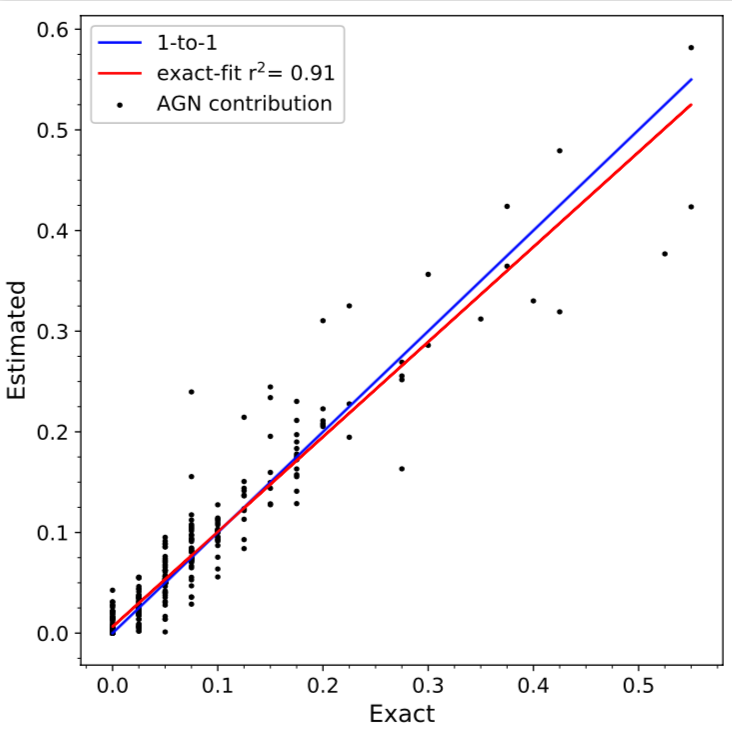}
    \caption{Comparsion of exact output AGN contribution from CIGALE and estimated AGN contribution from mock analysis of the spec-z sample. The blue line is the 1-to-1 relation line. The red line is the linear regression fit. The coefficient of determination is 0.91.}
    \label{fig:A1}
\end{figure}

\begin{figure}
    \centering
	\includegraphics[width=\linewidth]{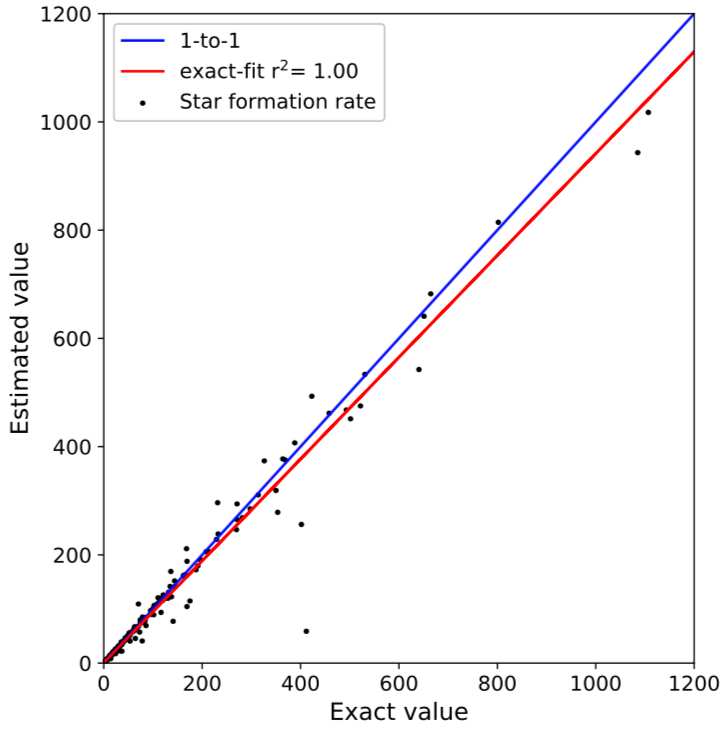}
    \caption{Comparsion of exact output SFR from CIGALE and estimated SFR from mock analysis of the spec-z sample. The blue line is the 1-to-1 relation line. The red line is the linear regression fit. The coefficient of determination is 1.00. In order to show the clear relation, two of the sources are not shown in the figure. The two sources are ID=133652 and ID=134015, with SFR= 1.0$^{+0.4}_{-0.4}$ $\times$ $10^{4}$ $M_{\odot} {\rm yr}^{-1}$ and SFR= 2.6$^{+0.2}_{-0.2}$ $\times$ $10^{3}$ $M_{\odot} {\rm yr}^{-1}$, respectively.}
    \label{fig:A2}
\end{figure}

Fig.~\ref{fig:A1} shows the comparsion of the original value of the AGN contribution from \begin{footnotesize}CIGALE\end{footnotesize} and estimated AGN contribution from mock analysis. The coefficient of determination is 0.91. Fig.~\ref{fig:A2} shows the comparsion of the original value of the SFR from \begin{footnotesize}CIGALE\end{footnotesize} and estimated SFR from mock analysis. The coefficient of determination is 1.00. The mock analysis of the physical properties (i.e. AGN contribution, SFR) shows that our SED fitting results are reliable.

\bsp	
\label{lastpage}
\end{document}